\documentclass[11pt,a4paper]{article}
\pdfoutput=1

\usepackage{jcappub}
\usepackage{graphicx}
\usepackage{epsfig}
\usepackage{subfigure}
\usepackage{comment}
\usepackage{slashed}
\usepackage{soul}
\usepackage{tabularx}
\usepackage{physics}
\usepackage[normalem]{ulem}
\usepackage{afterpage}
\usepackage{placeins}

\usepackage{array}
\newcolumntype{C}[1]{>{\centering\let\newline\\\arraybackslash\hspace{0pt}}m{#1}}

\newcommand{\be}{\begin{equation}} 
\newcommand{\ee}{\end{equation}}
\newcommand{\bea}{\begin{equation}\begin{aligned}} 
\newcommand{\eea}{\end{aligned}\end{equation}}
\newcommand{\ber}{\begin{eqnarray}}
\newcommand{\ear}{\end{eqnarray}}

\def\lsim{\mathrel{\raise.3ex\hbox{$<$\kern-.75em\lower1ex\hbox{$\sim$}}}}
\def\gsim{\mathrel{\raise.3ex\hbox{$>$\kern-.75em\lower1ex\hbox{$\sim$}}}}

\renewcommand{\tr}{{\rm tr}\,}

\newcommand{\GeV}{{\rm GeV}}

\newcommand{\GHz}{{\rm GHz}}

\newcommand{\ie}{{\it i.e.}}
\newcommand{\eg}{{\it e.g.\,}}

\newcommand{\td}{{\rm d}}

\newcommand{\mpl}{M_{\rm P}}

\newcommand{\eps}{\epsilon}

\newcommand{\tphi}{\tilde{\phi}}
\newcommand{\tU}{\tilde{U}}
\newcommand{\tlt}{\tilde{t}}

\newcommand{\tk}{{\tilde{k}}}

\newcommand{\mth}{m_{\rm th}}
\newcommand{\mufull}{\lambda}

\newcommand{\pk}{\text{peak}}
\newcommand{\amp}{\text{amp}}

\begin{document}

\title{Tachyonic Preheating in Plateau Inflation}

\author{Eemeli Tomberg}
\author{and Hardi Veerm\"{a}e}
\affiliation{NICPB, R\"avala 10, 10143 Tallinn, Estonia}  

\emailAdd{eemeli.tomberg@kbfi.ee}
\emailAdd{hardi.veermae@cern.ch}

\abstract{
Plateau inflation is an experimentally consistent framework in which the scale of inflation can be kept relatively low. Close to the edge of the plateau, scalar perturbations are subject to a strong tachyonic instability. Tachyonic preheating is realized when, after inflation, the oscillating inflaton repeatedly re-enters the plateau. We develop the analytic theory of this process and expand the linear approach by including backreaction between the coherent background and growing perturbations. For a family of plateau models, the analytic predictions are confronted with numerical estimates. Our analysis shows that the inflaton fragments in a fraction of an $e$-fold in all examples supporting tachyonic preheating, generalizing the results of previous similar studies. In these scenarios, the scalar-to-tensor ratio is tiny, $r<10^{-7}$.
}

\maketitle

\section{Introduction}
\label{sec:intro}

Cosmic inflation can solve many outstanding issues of the Big Bang model and successfully predict the spectrum of primordial perturbations~\cite{Starobinsky:1980te, Starobinsky:1983zz, Guth:1980zm, Linde:1981mu, Albrecht:1982wi, Linde:1983gd, Lyth:1998xn, Planck:2018jri}. 
The latest cosmic microwave background (CMB) measurements by the Planck satellite~\cite{Planck:2018jri} set tight limits on primordial perturbations, especially on the spectral index $n_s$ and the tensor-to-scalar ratio $r$. The measurements of $n_s$ support plateau inflation, a class of models where the inflaton field potential has a long, almost flat section at large field values~\cite{Martin:2013nzq,Planck:2018jri}. Tensor modes are yet to be observed, thus only an upper bound $r_{0.002}<0.056$ exists for the tensor-to-scalar ratio~\cite{Planck:2018jri,BICEP2:2018kqh}. Plateau inflation can accommodate extremely low values of $r$ practically undetectable even in next-generation CMB experiments, which are projected to have sensitivities of the order $r \sim 0.001$~\cite{Matsumura:2013aja,Hazumi:2019lys,CORE:2017oje}. If the upper limit on $r$ keeps shrinking, plateau models become more and more favoured.

In the hot Big Bang model, the post-inflationary universe must transition into radiation domination. This transition process is dubbed reheating~\cite{Kofman:1994rk,Kofman:1997yn}, and, in typical models, it proceeds through the inflaton field oscillating around the minimum of its potential and producing relativistic particles through non-perturbative processes and perturbative decay. The non-perturbative process, called preheating, often proceeds through parametric resonance, exciting fields coupled to the inflaton by rapidly and periodically changing their masses~\cite{Kofman:1997yn}. Another efficient preheating mechanism is tachyonic preheating, where the squared mass of a perturbed field becomes negative, and perturbations are produced due to a tachyonic instability~\cite{Felder:2000hj,Felder:2001kt}.

Although tachyonic preheating is often studied in hilltop~\cite{Felder:2000hj,Felder:2001kt,Garcia-Bellido:2001dqy} or hybrid inflation~\cite{Copeland:2002ku,Barnaby:2006cq,Barnaby:2006km,Garcia-Bellido:2007fiu,Dufaux:2008dn}, or in other multi-field setups~\cite{Greene:1997ge,Dufaux:2006ee,Abolhasani:2009nb,Battefeld:2009xw},
it is also generic to plateau inflation. In particular, the second derivative of the inflaton potential, \ie, the effective mass squared of perturbations, becomes negative at the plateau's edge, causing fragmentation of the homogeneous inflaton condensate\footnote{A similar effect, dubbed the flapping resonance, has been studied in the context of axion physics~\cite{Kitajima:2018zco,Fukunaga:2019unq}.}. This process is most efficient if the inflaton repeatedly returns to the plateau's edge during the oscillatory phase instead of being quickly damped by Hubble friction. As shown in previous studies of Palatini Higgs and $R^2$ inflation~\cite{Rubio:2019ypq,Karam:2020rpa,Karam:2021sno}, this happens when the plateau is extremely flat, the tensor-to-scalar ratio is small, $r\lesssim 10^{-7}$, and the energy scale of inflation is relatively low, with a Hubble parameter $H\lesssim 10^{10}$ GeV. In this regime, tachyonic preheating is, quite generally, the prevalent mechanism for particle production. Fragmentation is very quick---the field can fragment entirely in much less than an $e$-fold of expansion. This conclusion is backed up by lattice computations~\cite{Lozanov:2016hid, Lozanov:2017hjm, Krajewski:2018moi, Lozanov:2019ylm, Bhoonah:2020oov}. A better understanding of the preheating phase can improve the CMB predictions of these models, predict gravitational wave (GW) signals, and inspire further model building. For example, tachyonic preheating has been considered in the context of superheavy dark matter~\cite{Karam:2020rpa}.

In this paper, we study preheating in general models of inflation from an exponentially flat plateau. Our primary focus is the tachyonic fragmentation process, for which we outline the necessary conditions for fragmentation to be efficient. To reach this goal, we apply linear perturbation theory and extend it by including backreaction between perturbation growth and an adiabatically evolving background. In this way, we can probe the boundaries of the semianalytic linearized approach without resorting to full non-linear lattice simulations. For the inflationary models considered here, we construct simple (semi)analytic approximations for both the background oscillation dynamics and the perturbation spectrum. We do not explicitly consider inflaton's perturbative decay into Standard Model particles.

We work out preheating in detail for inflaton potentials of the form $U(\phi) = U_0 \tanh^n \phi/\phi_0$ and derive the preheating duration, the growth rate of the perturbations, and the wavenumber of the leading perturbation in terms of the model parameters. Consistently with previous studies~\cite{Rubio:2019ypq,Karam:2020rpa,Karam:2021sno}, we find that preheating is very efficient. Our results turn out to depend only weakly on the specifics of the model, which hints towards the existence of universal features underlying preheating in plateau inflation with exponentially flat potentials. 

The paper is organized as follows. In section \ref{sec:classical}, we outline the general features of plateau inflation and consider the evolution of the classical background in the inflationary and the oscillatory epochs. Section~\ref{sec:preheating} is dedicated to preheating. We give a general theoretical overview of the tachyonic instability of linear perturbations and numerically and analytically estimate how the growth of perturbations backreacts on the classical background. We conclude in section \ref{sec:concl}. Some technical details are gathered in the appendices. We use natural units $\hbar = c = \mpl = 1$ and the metric signature $(-,\!+,\!+,\!+)$.

\section{Plateau inflation}
\label{sec:classical}

We consider single-field inflation with a minimally coupled inflaton field $\phi$ given by the action
\be\label{action1}
    S = \int\td^4 x \sqrt{-g} \left[ \frac{1}{2} R - \frac{1}{2} (\partial \phi)^2 - U(\phi) \right] \, ,
\ee
%
where $g$ is the determinant of the metric $g_{\mu\nu}$, $R$ is the Ricci scalar, and $U$ is the inflaton potential. In the spatially flat Friedmann--Robertson--Walker Universe, expansion is controlled by the Friedmann equation $3 H^2 = \rho$, where $H\equiv \dot{a}/a$ is the Hubble parameter, $a$ is the scale factor, and dot indicates a derivative with respect to time. The energy density and pressure of the inflaton field are $\rho =  \dot\phi^2/2 + U$ and  $P = \dot\phi^2/2 - U$, respectively,
and the field equation reads
\be\label{FRW}
    \ddot\phi + 3 H \dot\phi +  U'(\phi) = 0 \, . 
\ee
In the models considered here, the evolution of the background can be split into two epochs: inflation and a subsequent phase in which the inflaton oscillates around the minimum of its potential. We will consider these separately in the following sections.

During the oscillatory phase, perturbations are generated from instabilities. We are interested in tachyonic preheating, which is triggered by the tachyonic instability. The existence of the latter sets requirements for the shape of the potential:
\begin{itemize}
    \item The potential must have a region in which 
    \be\label{cond1}
        m_{\rm eff}^2 \equiv U''(\phi) < 0 \, , 
    \ee
    thus making tachyonic instability possible. This condition is automatically satisfied if the potential has a plateau curving down towards a minimum. In that case, the tachyonic instability is strongest at the plateau's edge.
    
    \item The inflaton must repeatedly return to this unstable region, \eg, to the plateau. This is possible when Hubble friction is not too effective in damping the amplitude of the post-inflationary oscillations. In particular, inflation should end before reaching the edge of the plateau. This implies that the period of oscillations must be shorter than the Hubble time, which can be recast into the necessary condition
    \be\label{cond2}
        |U''| \gg H^2
    \ee
    somewhere in the oscillatory region. Furthermore, $U''$ is relevant for another timescale -- namely, it sets the timescale of the tachyonic instability. This should also be shorter than the Hubble scale, giving $-U'' \gg H^2$ again, so that the tachyonic instability can dominate over Hubble friction resulting in rapid mode growth. In our case, \eqref{cond2} must hold at the edge of the plateau.
\end{itemize}

Although we aim to address tachyonic preheating in general terms, for the sake of concreteness, we will work out numerical details for the potentials
\be\label{eq:U}
    U = U_{0} \tanh^n (\phi/\phi_0) \, ,
\ee
with $n$ an even integer. Such Einstein frame potentials arise naturally, \eg, in inflationary models in the Palatini formulation~\cite{Bauer:2008zj, Rubio:2018ogq, Enckell:2018hmo, Antoniadis:2018ywb, Tenkanen:2020dge} 
and the string theory motivated $T$-model $\alpha$-attractors~\cite{Ferrara:2013rsa, Kallosh:2013yoa, Carrasco:2015rva, Galante:2014ifa, Kallosh:2013hoa}. In the context of tachyonic preheating, the cases $n=2,3,4,6$ have been studied using linear perturbation theory~\cite{Rubio:2019ypq,Karam:2021sno,Lozanov:2016hid,Lozanov:2017hjm} for a wide range of $\phi_0$ values, and using a non-linear lattice setup~\cite{Lozanov:2016hid,Lozanov:2017hjm,Lozanov:2019ylm} with $\phi_0 \gtrsim 10^{-2}$ (corresponding to the edge of tachyonicity, see below). In addition, production of GWs during preheating for $n=2$ and $\phi_0 \gtrsim 10^{-4}$ has been studied on a lattice~\cite{Bhoonah:2020oov}. Analytical conditions for inflaton fragmentation in similar models were previously studied in \cite{Kim:2017duj,Lloyd-Stubbs:2020pvx}. We aim to generalize these studies by providing a systematic understanding of the linear preheating regime and deriving scaling relations applicable for a wide range of $\phi_0$ and $n$.

The potentials \eqref{eq:U} have a plateau when $\phi \gtrsim \phi_0$ and the condition~\eqref{cond1} is satisfied on this plateau. For future convenience, we define the mass scale
\be\label{eq:mth}
    \mth \equiv \frac{\sqrt{U_{0}}}{\phi_0} \, ,
\ee
characterizing the magnitude of the effective mass of scalar perturbations, $m^2_{\rm eff} = U''$. For example, at the edge of the plateau, the effective squared mass of scalar fluctuations reaches values as low as $U''_{\rm min} \approx - 0.65 \mth^2$ almost independently of $n$. Thus, $\mth$ also controls the strength of the tachyonic instability. The second condition \eqref{cond2} then implies that the tachyonic instability can be active when
\be\label{cond2b}
    \phi_0 \ll 1 \, ,
\ee
where we used that, close to the plateau, inflaton's energy density will be potential dominated, \ie, $H^2 \approx U_0/3$, even in the oscillatory regime.

Interestingly, as we will see below, many of our models' features, from inflationary phenomenology to the specifics of tachyonic preheating, can be deduced from the large $\phi$ behaviour,
\be\label{eq:expU}
    U \sim U_0 (1 - A \exp(-2\phi/\phi_0)) \, ,    \qquad \qquad \phi \gtrsim \phi_0 \, ,
\ee
where $A = 2 n$ for the potentials \eqref{eq:U}. This can be understood intuitively by noting that the oscillating field does not spend much time around the minimum of the potential, so both inflationary dynamics and tachyonic preheating are determined by features of the plateau. We expect the large-$\phi$ behaviour to be a deciding factor also for other similar classes of potentials. Such models are fairly common~\cite{Martin:2013tda, Martin:2013nzq} and include, \eg, the well-known Starobinsky inflation~\cite{Starobinsky:1980te,Starobinsky:1983zz} and the metric Higgs inflation~\cite{Bezrukov:2007ep, Bezrukov:2013fka, Rubio:2018ogq}, though for them $\phi_0 = \sqrt{6}$, which lies outside of the tachyonic parameter region.

\subsection{Inflationary phase}

Plateau inflation can be accurately studied with the slow-roll approximation. Using the potential slow-roll parameters $\epsilon_U \equiv \frac{1}{2}(U'/U)^2$ and $\eta_U \equiv U''/U$, we can write down the inflationary CMB observables: the power spectrum of scalar perturbations, the spectral index, and the tensor-to-scalar ratio,
\be
    A_s = \frac{1}{24 \pi^2} \frac{U}{\epsilon_U}, \qquad
    n_s = 1 - 6 \epsilon_U + 2 \eta_U, \qquad
    r = 16 \epsilon_U \, .
\ee
All quantities are computed when the comoving pivot scale $k_* = 0.05 \text{Mpc}^{-1}$ exits the horizon.

For the class of potentials \eqref{eq:U}, the number of $e$-folds of inflationary expansion corresponding to the field value $\phi$ is
\be
    N \approx \int^{\phi}_{0} \frac{\td \phi}{\sqrt{\eps_U}} =  \frac{\phi_0^2}{2n}\sinh^2\left(\frac{\phi}{\phi_0}\right)
    \qquad \Rightarrow \qquad 
    \phi 
    \approx \frac{\phi_0}{2} \ln \frac{8nN}{\phi_0^2} \, ,
\ee
where the last identity holds when $N \gg 1$.\footnote{To estimate the number of $e$-folds we begin integration from $\phi = 0$. This introduces a negligible error. Note also that plateau inflation can end violating the second slow-roll condition $|\eta_U| \ll 1$ instead of the first one $|\epsilon_U| \ll 1$.} Thus, at the leading order in the large $N$ approximation,
\be\label{eq:CMB_predictions}
    A_s \approx \frac{N^2 \mth^2}{3 \pi^2} 
    , \qquad
    n_s \approx 1 - \frac{2}{N}
    , \qquad
    r \approx \frac{2\phi_0^2}{N^2} 
    \, .
\ee
Note that the dependence on $n$ has dropped out. For these expressions to hold, we need $\phi_0^2 \ll 2 n N$, which, by Eq.~\eqref{cond2b}, is satisfied whenever we require tachyonic preheating to be active. The inflationary predictions \eqref{eq:CMB_predictions} are a generic feature of models in which the potential has the exponential behaviour \eqref{eq:expU}.

The number of $e$-folds $N_*$ corresponding to the CMB pivot scale depends on the details of reheating~\cite{Liddle:2003as}. Assuming instantaneous reheating, we have $N_* \approx 61 + \frac{1}{4}\ln(U_0)$. In the following we will neglect the $\ln(U_0)$ dependence and use  $N_* \approx 50$ yielding $n_s \approx 0.96$ compatible with the Planck measurement~\cite{Planck:2018jri}. The observed power spectrum strength $A_s = 2.1 \times 10^{-9}$ fixes the value of our mass parameter, 
\be\label{eq:m_value}
    \mth = 5\times 10^{-6} \, .
\ee
The condition for tachyonic preheating \eqref{cond2b} further implies that $r \ll 8 \times 10^{-4}$, $U_0 \ll 2.5 \times 10^{-11}$.

\subsection{Oscillatory phase}
\label{sec:oscillations}

In the oscillatory phase, we closely follow the formalism of Ref.~\cite{Karam:2021sno} and work in the adiabatic limit in which the oscillation period is much shorter than the Hubble scale. In this case, a single oscillation can be studied by neglecting the expansion. The damping of energy density over multiple oscillations, on the other hand, can be inferred from the continuity equation
\be\label{eq:cont}
    \dot\rho + 3H (\rho + \bar P) = 0\,,
\ee
by considering the time-averaged pressure $\bar P$ and an effective equation of state $\bar P = \bar P(\rho)$~\cite{Turner:1983he}. We remark that, by using \eqref{eq:cont}, we neglect the backreaction from perturbation growth. If the latter is extremely fast, the coherent field may fragment before completing a single oscillation, so an oscillatory regime is never realized. We will return to this issue in section~\ref{sec:results}.

In the adiabatic approximation, we first consider the evolution of the homogeneous field on a flat background. This can be described as a mechanical system with the action $\int \td t P(\phi,\dot\phi)$ and a conserved energy density $\rho(\phi,\dot\phi) = \dot\phi\partial P/\partial \dot\phi - P$. All relevant quantities can be computed from the abbreviated action  
\be\label{def:W}
    W(\rho) 
    \equiv \int^{\phi_2}_{\phi_1} \td \phi \left.\frac{\partial P}{\partial \dot \phi}\right|_{\dot\phi = \dot\phi(\phi,\rho)}
    = 2 \int^{\phi_\amp}_{0} \td \phi \sqrt{2(\rho - U(\phi))} \, ,
\ee
between the turning points $\phi_{i}(\rho)$ which are defined as values of the field at which the velocity vanishes, $\dot\phi(\phi_{i},\rho) = 0$. The expression $\dot\phi(\phi,\rho)$ is obtained by inverting $\rho = \dot\phi\partial P/\partial \dot\phi - P$. While the first expression in \eqref{def:W} is general, the second expression applies in the case of a canonical kinetic term, as in \eqref{action1}, and for symmetric potentials, \ie\, $U(\phi) = U(-\phi)$, which we assume throughout the paper unless specified otherwise. Then $|\phi_{i}| \equiv \phi_\amp$, the oscillation amplitude of the coherent field component.

The half-period and the time-averaged pressure are~\cite{Karam:2021sno} 
\be\label{eq:T_and_P}
    T = \partial_{\rho} W,    \qquad\qquad
    \bar P = W/T - \rho \, .
\ee
By eliminating the time-averaged pressure from the continuity equation~\eqref{eq:cont}, we obtain a closed form equation for the energy density,
\be\label{eq:rho_friction_evolution}
    \dot{\rho} + 3H W/W'(\rho) = 0 \, ,
\ee
where the second term describes the average effect of Hubble friction for an oscillating field in a general potential. Note that the abbreviated action $W$ obeys the considerably simpler equation $\dot W + 3 H W = 0$ and is thus conserved in a comoving volume, that is, $W \propto a^{-3}$. The fraction of energy lost during a half-oscillation is
\be\label{eq:delta_rho}
    \frac{\Delta \rho}{\rho} 
    = -\frac{W}{H} \, ,
\ee
given the scalar dominates the energy density so that $3H^2=\rho$.

\begin{figure}
\begin{center}
	\includegraphics[width=.95\linewidth]{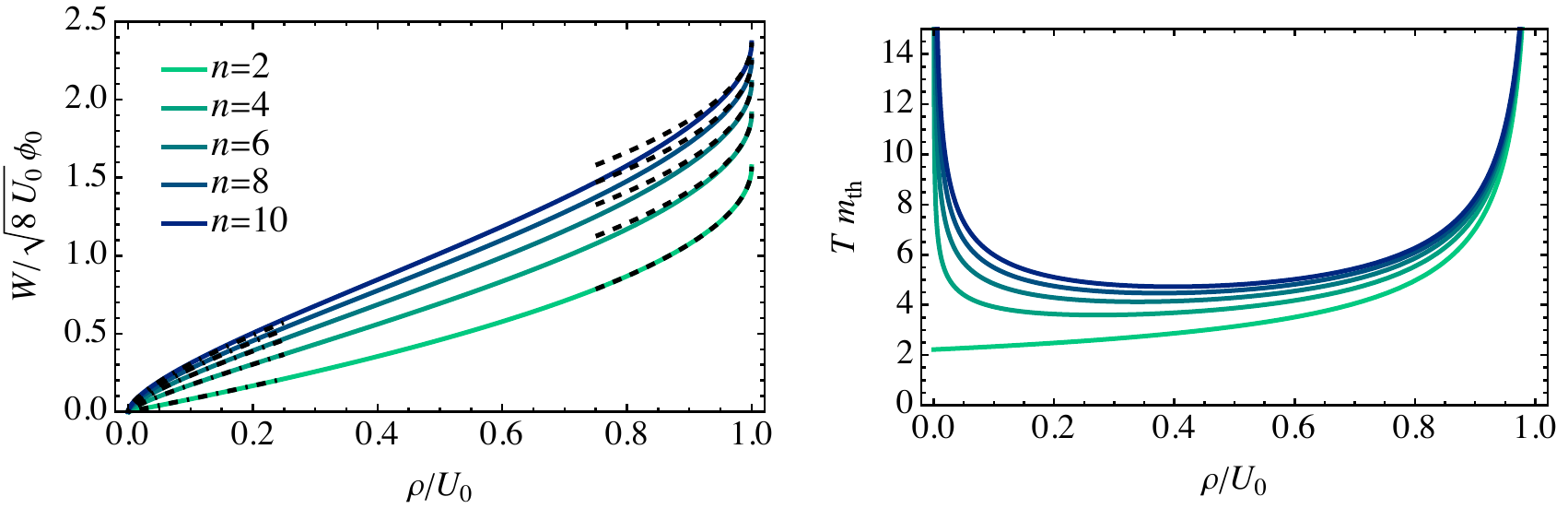}
\end{center}
\vspace{-7mm}
\caption{\emph{Left panel:} The abbreviated action $W$ (solid) for the potential \eqref{eq:U} with $n = 2,4,6,8,10$ together with its asymptotics at $\rho\sim 0$ (dot-dashed) and $\rho\sim U_0$ (dashed). \emph{Right panel:} The dependence of the half-period on the energy density.}
\label{fig:W}
\end{figure}

Having these theoretical tools at our disposal, let us consider the potential \eqref{eq:U}. The abbreviated action corresponding to \eqref{eq:U} is given by
\be\label{eq:W}
    W \approx \sqrt{8 U_0}\phi_0 f_n(\rho/U_0) \, , 
    \qquad
    f_n(x) \equiv x^{\frac{1}{n}+\frac{1}{2}}\int^{1}_{0} \frac{\sqrt{1-y^n}}{1-x^{2/n} y^2} \td y \, .
\ee
Close to the plateau ($\rho \approx U_0$) and the minimum ($\rho \ll U_0$) we find
\be\label{eq:fn}
    f_n(x) \sim
\left\{\begin{array}{lc}
     \frac{\sqrt{\pi}}{2n}x^{\frac{1}{n}+\frac{1}{2}} \left( \frac{\Gamma(1/n)}{\Gamma(3/2+1/n)} + \frac{\Gamma(3/n)/\Gamma(3/2+3/n)}{x^{-\frac{2}{n}} - \frac{\Gamma(3/2+3/n) \Gamma(5/n)}{\Gamma(3/2+5/n) \Gamma(3/n)} } \right)&,\mbox{ when }  x \sim 0\\
     f_n(1) - \frac{\pi}{2} \sqrt{1 - x} &,\mbox{ when } x\sim 1 \, .
\end{array}\right. 
\ee
The abbreviated action together with these asymptotics is shown in Fig.~\ref{fig:W} for selected values of $n$. When $n=2$, the $x\sim 1$ asymptotic matches $f_2(x)$ exactly (see appendix \ref{app:n=2}).  For higher $n$, analytic expressions can be obtained in terms of hypergeometric functions.\footnote{For even $n$,
\begin{equation*}\begin{aligned}
    f_n(x) 
&    = x^{\frac{1}{n}+\frac{1}{2}}\int^{1}_{0} \sqrt{1-y^n} \sum^{n/2-1}_{k=0}\sum^{\infty}_{m=0}(x^{\frac{2}{n}} y^2)^{\frac{n}{2} m + k} \td y \\
&    = \frac{\sqrt{\pi}}{2n} \sum^{n/2-1}_{k=0} x^{\frac{2k+1}{n} + \frac{1}{2}} \frac{\Gamma\left(\frac{1+2k}{n}\right)}{\Gamma\left(\frac{1+2k}{n}+\frac{3}{2}\right)} {}_2F_{1}\left(1,\frac{1+2k}{n},\frac{1+2k}{n} + \frac{3}{2};x\right).
\end{aligned}\end{equation*}
} As a consistency check, we find that the leading term of the $x\sim 0$ asymptotic matches the corresponding result for $\phi^n$ potentials~\cite{Turner:1983he}.

Consider now the first oscillations, as they tend to be the most relevant for the dynamics of tachyonic preheating. For the tachyonic instability to become active, the first oscillations must return to the plateau, so they take place in the regime in which $\rho \approx U_0$.
From \eqref{eq:delta_rho}, 
\be
    \frac{\Delta \rho}{\rho} 
    \approx \phi_0 2\sqrt{6} f_n(1) \, ,
\ee
so the energy loss per oscillation from Hubble friction is indeed small when the condition \eqref{cond2b} holds, \ie, when $\phi_0 \ll 1$. At the first turning point, $\rho = U_0 - \Delta \rho = U(\phi_{\amp,i})$, so the initial oscillation amplitude is
\be\label{eq:phi_i}
    \phi_{\amp,i}
    \approx -\frac{\phi_0}{2} \ln \left( \phi_0 \frac{\sqrt{6} f_n(1)}{n} \right) \, ,
\ee
so the initial value of the field lies indeed on the plateau ($\phi_{\amp,i} \gtrsim \phi_0$) as long as $\phi_0$ is small.
As a consistency check, the slow-roll parameters at this point are $\epsilon_U \approx 48f_n(1)$ and $\eta_U \approx -8\sqrt{6}f_n(1)/\phi_0$. Both are large, so oscillations start after the slow-roll epoch. Specifically, $\eta_U$ can be several orders above one at that point.

By \eqref{eq:T_and_P} and \eqref{eq:fn}, the half-period of the first oscillations reads
\be\label{eq:T}
    T 
    \approx \frac{\pi \phi_0}{\sqrt{2(U_0 - \rho )}}
    = \frac{\pi \mth^{-1}}{\sqrt{2(1 - \rho/U_0)}} \, .
\ee
Half-periods for the next four even $n$ are shown in Fig.~\ref{fig:W}. The half-period is proportional to the inverse mass scale $\mth^{-1}$ but diverges for large oscillation amplitudes when $\rho \to U_0$, so the condition \eqref{cond2} is necessary but might not be sufficient to guarantee oscillations faster than the Hubble rate. However, fixing the initial oscillation amplitude from \eqref{eq:phi_i} shows that the initial half-period is
\be\label{eq:T_i}
    H T_{i} \approx \frac{0.6\sqrt{\phi_0}}{\sqrt{f_n(1)}} \, ,
\ee 
so the adiabaticity condition $T \ll 1/H$ still applies as long as $\phi_0 \ll 1$, consistent with \eqref{cond2b}. Plugging in $H\sim\sqrt{U_0}$, we further see that $T_i \sim (\mth \sqrt{\phi_0})^{-1}$.

Since $W \propto a^{-3}$, Eq.~\eqref{eq:W} implies that for oscillations reaching the plateau, $\rho\approx U_0$, the background's energy density transitions from vacuum energy domination as $\rho/U_0 \propto 1 - (a^{-3} - c_n)^2$, where $c_n$ is a numerical constant depending on $n$. A special case of such behaviour was observed in Ref.~\cite{Karam:2021sno}. For oscillations at the bottom of the potential, $\rho \ll U_0$, we recover the well-known dilution $\rho \propto a^{-6n/(2+n)}$ of $\phi^{n}$ potentials~\cite{Turner:1983he}. However, the latter scaling is not realized in our models due to the fast fragmentation of the homogeneous field.

Finally, eqs.~\eqref{eq:W} and ~\eqref{eq:T} show that the half-period in the $\rho \approx U_0$ region is independent of $n$ in the first approximation, and the abbreviated action depends on $n$ only through a numerical constant (see Fig.~\ref{fig:W} for specific values of $n$). This behaviour is a general consequence of the exponential large field behaviour \eqref{eq:expU}. Indeed, assuming a symmetric potential with an asymptotic exponential tail of the form \eqref{eq:expU}, a straightforward computation shows the half-period is approximately \eqref{eq:T} since most of the time is spent on the plateau. Integrating Eq.~\eqref{eq:T_and_P} then gives the general shape
\be\label{eq:W_expU}
    W \sim \pi \sqrt{2U_0} \phi_0  (C - \sqrt{1 - \rho/U_0}), \qquad \mbox{when} \qquad \rho \sim U_0 \, ,
\ee
where $C$ is a number that depends on the shape of the potential away from the asymptotic region, \eg, for the potential considered here with $n = 2,4,6,8,10$ the corresponding values read $C = 1, 1.22, 1.34, 1.44, 1.51$, respectively. Thus, the above results for the $U_0 \approx \rho$ case also apply for this more general class of potentials to which \eqref{eq:U} belongs.

\section{Preheating} 
\label{sec:preheating}

While the homogeneous inflaton condensate oscillates, its perturbations start to grow. This growth proceeds through two channels: parametric resonance~\cite{Traschen:1990sw, Kofman:1994rk, Shtanov:1994ce, Kofman:1997yn} and tachyonic instability~\cite{Felder:2000hj,Felder:2001kt}. Both effects can lead to preheating, \ie, the complete fragmentation of the coherent background field. Parametric resonance takes place due to the time dependence of the frequency of perturbations $\omega_k$ and can excite modes in frequency bands corresponding to integer multipliers of a base frequency. We focus, instead, on scenarios in which the tachyonic instability, triggered by a negative squared frequency $\omega_k^2<0$, dominates preheating. 

The scalar perturbations $\delta\phi_k$ with wavenumber $k$ follow, to linear order, the equations of motion
\be\label{eq:deltaphi_eq}
    \delta\ddot{\phi}_k + 3H\delta\dot{\phi}_k + \omega_k^2\delta\phi_k = 0 \, ,
    \qquad
    \omega_k^2 \equiv k^2/a^2 + U''(\phi) \, ,
\ee
together with the Bunch--Davies initial conditions, $\delta\phi_k = 1/(\sqrt{2k}a)$, $\delta\dot{\phi}_k = -i(k/a)\delta\phi_k$~\cite{Birrell:1982ix} (see appendix~\ref{sec:mode_evol}). The problem then contains two relevant timescales: the half-period of the background oscillations $T$ sets the timescale for the time variation of $\omega_k$, while $\omega_k$ itself sets the relevant timescale for the growth of the perturbations $\delta\phi_k$. As we will show, these two scales are of similar order.

We will next discuss perturbation growth and the solutions of \eqref{eq:deltaphi_eq} generally. Focusing on our potentials \eqref{eq:U}, we will then provide simple analytical fits for the fastest growing modes. Continuing the approach of section \ref{sec:oscillations}, we will treat the effect of Hubble friction adiabatically. Assuming the field does not fragment too rapidly, we will also estimate how perturbation growth backreacts on the evolution of the background field. In particular, we compute the duration of the preheating process.

\subsection{Adiabatic mode growth} 
\label{sec:mode}

\begin{figure}
    \centering
    \hspace{-5mm}
    \includegraphics[width=.99\linewidth]{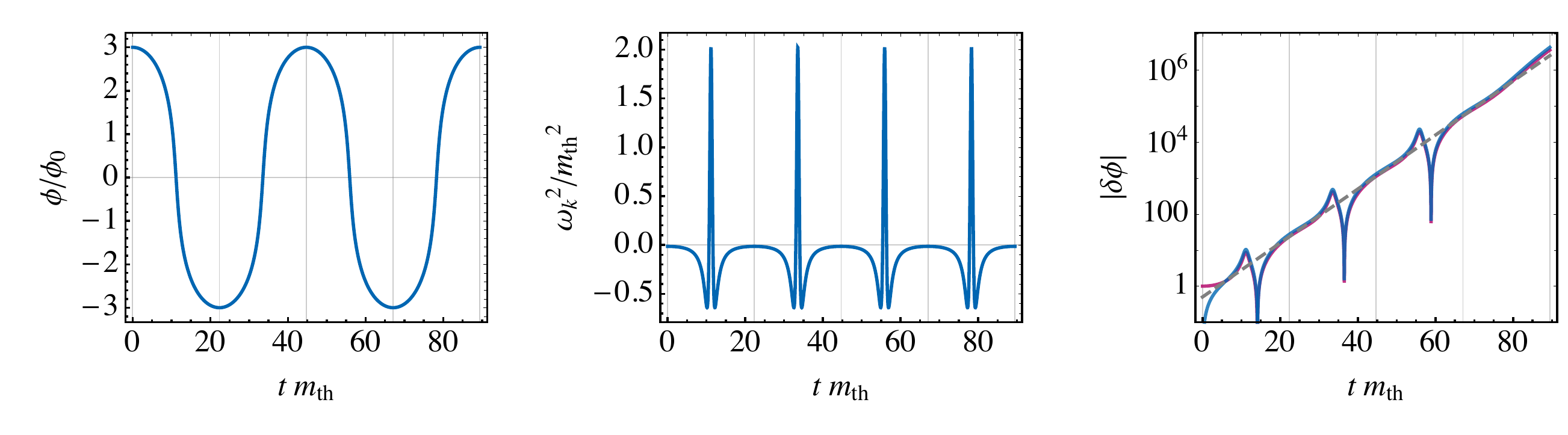}
    \vspace{-5mm}
    \caption{
    \emph{Left:} Two oscillations of the background field in the potential \eqref{eq:U} with $n=2$ and an amplitude $\phi_\amp=3\phi_0$ in the $H \to 0$ limit. 
    \emph{Middle:} Evolution of $\omega_k$ for the peak mode with $k=k_\pk$ during the same period.
    \emph{Right:} Independent solutions for the peak mode (red and blue) and the average exponential growth $e^{\mu_k t}$ (dashed grey). 
    }
   \label{fig:mode}
\end{figure}

In plateau inflation, the characteristic timescales for the perturbations are much shorter than the Hubble time and the effect of expansion can be studied perturbatively as we did for the classical background. Thus, at the leading order, mode growth is computed in the $H \to 0$ limit. In this case, Eq.~\eqref{eq:deltaphi_eq} simplifies to the Hill equation
\be\label{eq:deltaphi_eq_noH}
    \delta\ddot{\phi}_k + \omega_k^2\delta\phi_k \approx 0 \, ,
\ee
where the period of $\omega_k^2$ matches the background's half-period $T$.\footnote{This holds for symmetric potentials assumed here. If the potential is asymmetric, then the period of $\omega_k^2$ coincides with the full period of background oscillations instead of the half-period.} This is illustrated in Fig.~\ref{fig:mode}. Floquet's theorem then invites us to seek solutions that are periodic up to a growth factor~\cite{1968ZaMM...48R.138R}, \ie, 
\be\label{eq:muT_def}
    \delta\phi_k(t+T) = e^{\mufull_k T}\, \delta\phi_k(t) \, ,
\ee
where $\mufull_k$ is a Floquet exponent. Since \eqref{eq:deltaphi_eq_noH} is a second order equation, there will be two Floquet exponents. The growth rate of the mode is given by the largest real part of these exponents,
\be\label{def:mu_k}
    \mu_k \equiv \max \Re \mufull_{k}.
\ee
In this subsection, we will focus on the computation of this growth rate. 

Given any two independent solutions $u_1(t)$ and $u_2(t)$ of the $H \to 0$ mode equation \eqref{eq:deltaphi_eq_noH}, their Wronskian matrix $w(t)$ evolves as
\be\label{eq:G}
    w(t+T) = w(t) G \, , \qquad
    w(t) \equiv
\begin{pmatrix}
    u_1(t) & u_2(t) \\
    \dot{u}_1(t) &  \dot{u}_2(t) \\   
\end{pmatrix},
\ee
where $G \equiv w(0)^{-1} w(T)$ is a constant\footnote{One can show that $\td (w(t)^{-1} w(t+T))/\td t = w(t)^{-1} (\dot{w}(t+T)-\dot{w}(t) w(t)^{-1} w(t+T)) = 0$ by noting that the mode equation \eqref{eq:deltaphi_eq_noH} implies that the Wronskian matrix satisfies $\dot w = D w$ where $D(t+T) = \pm D(t)$ is a periodic matrix.} matrix dubbed the monodromy matrix. Eigenfunctions of $G$ satsify \eqref{eq:muT_def} and thus $e^{\mufull_k T}$ must be an eigenvalue of $G$. The most convenient choice of $u_{1,2}$ depends on the context.\footnote{For different choices of the two independent solutions $u_{1,2}$, the corresponding $G$ matrices are related by a similarity transformation $G \to L G L^{-1}$, thus the Floquet exponents are independent of the choice of $u_{1,2}$.} For example, when looking for the Floquet exponent numerically, we solve \eqref{eq:muT_def} with initial conditions $u_1 = \dot{u}_2 = 1$, $u_2 = \dot{u}_1 = 0$, or equivalently, $w(0) = 1$, so that $G = w(T)$.

\begin{figure}
\begin{center}
    \hspace{-5mm}
	\includegraphics[width=.94\linewidth]{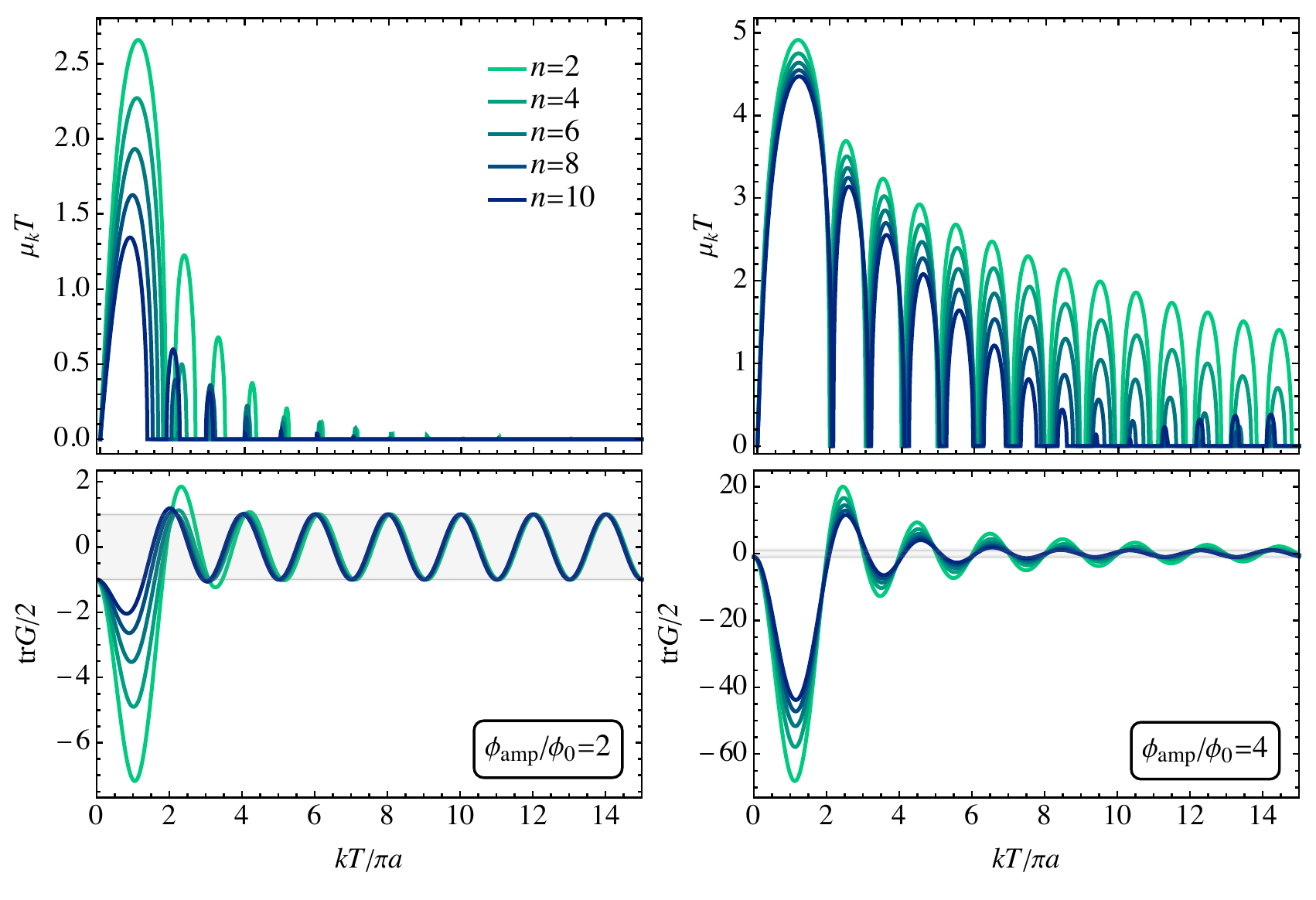}
\end{center}
\vspace{-5mm}
\caption{\emph{Upper panels:} Growth rate $\mu_k$ of modes in the $H \to 0$ limit for the potential \eqref{eq:U} with $n=2,4,6,8,10$ assuming a background with amplitude $\phi_{\rm amp} = 2 \phi_0$ (left panels) and $\phi_{\rm amp} = 4 \phi_0$ (right panels). \emph{Lower panels:} $\tr G/2$ corresponding to the panel above it. The grey band shows the region in which the modes are stable, \ie, $\mu_k=0$.}
\label{fig:muk_U}
\end{figure}

The equations of motion imply that the Wronskian $\det w(t)$ is constant in time, so taking the determinant of both sides of \eqref{eq:G} we find that $\det G = 1$. Therefore, $\mufull_{k,-} = -\mufull_{k,+}$, and, since $G$ is a real matrix, $\mufull_{k\pm}$ are either real or purely imaginary. Both Floquet exponents are given by the simple expression
\be\label{eq:mu_TrG}
    \mufull_{k,\pm} = \pm T^{-1}{\rm acosh}\left( \frac{1}{2} \tr G \right).
\ee
We drop the $\pm$ subindices and denote by $\mufull_k$ the eigenvalue with positive real part $\mu_k = \Re \mufull_k \geq 0$. The Floquet exponents are real (imaginary) if $|\tr G| > 2$ ($|\tr G| < 2$). If $\mufull_k$ is real, one of the solutions is growing and quickly starts to dominate over the other solution, which is decaying. Such growth marks an instability of the mode.

An example of mode growth in the $H \to 0$ limit is shown in Fig.~\ref{fig:mode}. Through $\omega_k^2$, the growth rate $\mu_k$ depends on the shape of the potential and the oscillation amplitude, in addition to the wavenumber $k$. Fig.~\ref{fig:mode} uses the potential \eqref{eq:U} with $n=2$ where the background field, depicted in the left panel, oscillates with an amplitude $\phi_\amp=3\phi_0$ and a half-period $T = 22.4\mth^{-1}$. The right panel of Fig.~\ref{fig:mode} shows the evolution of two independent solutions to the mode equation \eqref{eq:deltaphi_eq_noH} for the fastest growing mode $k_{\pk}/a \approx 3.54/T$. Both of these solutions contain the growing mode, so they look visually identical after a first half-period. In the middle panel, the squared frequency of the mode $\omega_k^2$ is seen to take mostly negative values except when the background field briefly crosses the origin. Thus, most of the time, the mode follows well the simple exponential curve $e^{\mu_k t}$, shown by a dashed grey line in the left panel. Such exponential growth would be exact when $\omega_k^2$ is constant and negative.

Numerical estimates of the growth rate $\mu_k$ over a range of $k$-values are shown in Fig.~\ref{fig:muk_U}, together with the corresponding trace of the monodromy matrix (Hill's discriminant), for potentials \eqref{eq:U} with $n=2,4,6,8,10$ 
and a fixed background amplitude $\phi_\amp$.\footnote{Floquet charts for $n=2,3,4,6$ have been computed also in Ref.~\cite{Lozanov:2016hid, Lozanov:2017hjm}.} As explained in appendix~\ref{sec:mode_evol}, dimensionless quantities such as $\mu_k T$ do not depend explicitly on dimensionful model parameters such as $\phi_0$ and $U_0$. Note that fixing $\phi_\amp/\phi_0$ means that both $\rho$ and $T$ vary with $n$. With this convention, the first peak only varies slightly for different $n$ when the amplitude is large, $\phi_\amp = 4 \phi_0$. The differences become more visible when $\phi_\amp = 2 \phi_0$, although the order of magnitude remains unchanged. The reduction of differences between different $n$ with increasing amplitudes is expected as for larger $\phi_\amp$ the background dynamics becomes more and more dominated by the $n$-independent exponential tail. In all, when $\phi_\amp \gg \phi_0$, the first and dominant tachyonic peak has a nearly universal shape. 

\begin{figure}
\begin{center}
    \hspace{-5mm}
	\includegraphics[width=.95\linewidth]{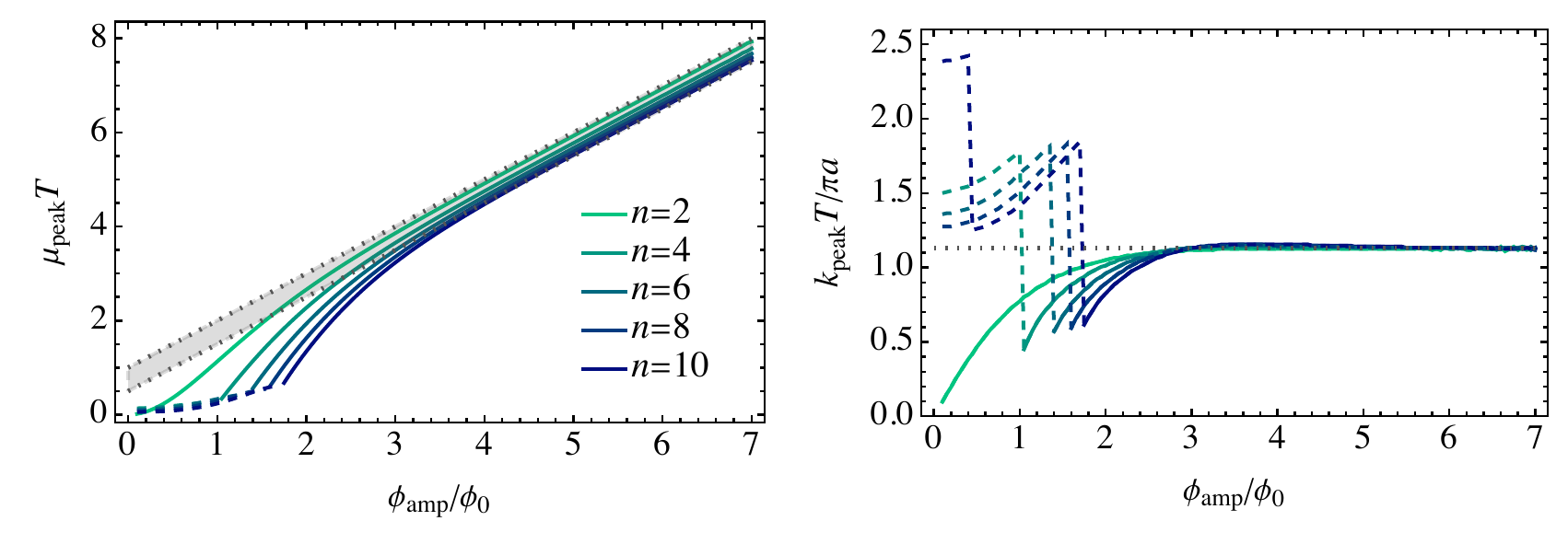}
\end{center}
\vspace{-5mm}
\caption{The dependence of maximal value of $\mu_k$ {(left panel)} and the wavenumber of the corresponding mode $k_\pk$ {(right panel)} on the oscillation amplitude $\phi_\amp$ for the potential \eqref{eq:U} with $n=2,4,6,8,10$. The dashed lines correspond to cases where the $\mu_k$ does not peak in the first instability band but in one the secondary bands instead. The grey dotted lines show the asymptotic behaviour at large amplitudes: $\mu_\pk T - \phi_\amp/\phi_0 \in (1/2,1)$ in the left panel and $k_\pk T = 3.54$ in the right panel.}
\label{fig:peak}
\end{figure}

The same cannot be said for higher $k$ peaks---even when the first peaks are matched. The higher peaks can vary significantly for different $n$ (even for $\phi_\amp = 4 \phi_0$). However, as can be observed from Fig.~\ref{fig:muk_U}, the instability bands vary nearly periodically in $k$-space. For $\phi_\amp = 4 \phi_0$, the end of the instability bands is approximately $k/a \approx j\pi/T$, with $j\geq2$ an integer. The secondary instability bands arise predominantly due to parametric resonance rather than the tachyonic instability as $\omega_k^2$ is mostly positive in these bands. They are subleading compared to the dominant tachyonic peak.

The growth rate and wavenumber of the fastest growing mode are shown in Fig.~\ref{fig:peak}. For sufficiently large amplitudes they can be approximated by
\be\label{eq:appr_peak}
    \mu_\pk \approx \frac{\phi_\amp/\phi_0 + c}{T} \, , \qquad
    k_\pk/a \approx \frac{3.54}{T} \qquad \qquad
    (\phi_\amp \gtrsim 3\phi_0)\, ,
\ee
where $c$ is an $\mathcal{O}(1)$ coefficient that lies in the range (0.5,1) for the cases considered in Fig.~\ref{fig:peak}.\footnote{In detail, for $n=2,4,6,8,10$ we find $c=0.93,0.79,0.68,0.6,0.53$, respectively.} Since  $\phi_0$ is related to the maximal oscillation amplitude by Eq.~\eqref{eq:phi_i}, $\phi_0$ must be relatively small for these approximations to be valid. The condition $\phi_{\amp,i}/\phi_0 \gtrsim 3$ translates into $\phi_0 \lesssim 10^{-3}$. For $\phi_0 \approx 10^{-2}$, $\phi_{\amp,i}/\phi_0$ goes from 2 to 2.6 when $n$ varies from 2 to 10, so the above approximation barely holds even during the initial oscillations, although it still gives the correct order of magnitude for $\mu_\pk$ and $k_\pk$. For even larger $\phi_0$ and thus for lower initial amplitudes $\phi_{\amp,i}$, the instability within the first band weakens and the secondary peaks begin to dominate indicating that tachyonicity becomes less relevant for preheating.  

Relying on \eqref{eq:appr_peak}, we can derive simple order-of-magnitude estimates in terms of the model parameters. $T_i \sim \mth^{-1} \phi_0^{-1/2}$ gives $k_\pk\sim\mu_\pk\sim \mth \sqrt{\phi_0}$. This is also the order of magnitude of $\sqrt{-U''(\phi_{\amp,i})}$, the scale of tachyonicity, so all relevant time and energy scales are of similar order.

We remark that the half-period $T$ grows rapidly when $\phi_\amp \to \infty$. Thus, by Eq.~\eqref{eq:appr_peak}, $\mu_\pk$ will generally decrease in this limit. Generally, $\mu_\pk$ will peak when $\phi_\amp \approx \mathcal{O}(1)$; the exact value depends on $n$. When the amplitude decreases below this value as the field relaxes to the minimum, $\mu_\pk$ will approach 0.

According to the numerical estimates in Figs.~\ref{fig:muk_U} and~\ref{fig:peak}, $\mu_\pk \gtrsim T^{-1}$ in the tachyonic regime, that is, the fastest growing mode can increase significantly during a half-oscillation. Therefore, after these modes begin to contribute noticeably to the total energy density, we expect that strong feedback effects will lead to a fast breakdown of adiabatic background evolution. Moreover, since the instability gets stronger with decreasing $\phi_0$, the field may become significantly fragmented even before the first oscillation is complete, and thus, it is expected that the adiabatic approach breaks down for very small $\phi_0$. This will be shown in section \ref{sec:results}.

Finally, consider the asymptotics of $\mu_k$. Firstly, the $k=0$ mode $\delta\phi_0$ does not grow, \eg, in Fig.~\ref{fig:muk_U}, we see that $\mu_k \to 0$ as $k \to 0$. Specifically, the $k \to 0$ asymptotic for symmetric\footnote{In case of asymmetric potentials, one must replace $T$ with the full period, $W$ with the corresponding integral over the full period and flip the sign of the monodromy matrix, \ie, $\tr G \to -\tr G$.} potentials is (for a derivation, see appendix \ref{sec:mu_asymp})
\be\label{eq:k0growth}
    \mufull_k = -\frac{i\pi}{T} + (k/a) \frac{\sqrt{W \partial_\rho T}} {T} + \mathcal{O}(k/a)^3
    \quad \Leftrightarrow \quad
    \frac{1}{2}\tr G = -1 - \frac{(k/a)^2}{2} W \partial_\rho T + \mathcal{O}(k/a)^4 \, .
\ee
Thus the position of the first stability band is tied to whether the period of background oscillations decreases with decreasing background energy density or not: as long as $\partial_\rho T>0$, the first instability band begins at $k=0$ since $\tr G < -2$ for modes surrounding $k=0$. In particular, the slope of $T$ shown in Fig.~\eqref{fig:W} implies that the first instability band starts at $k=0$ for arbitrarily small background energy densities only when $n=2$ -- for larger $n$, however, the infrared modes are stable when the oscillations are sufficiently damped. In general, for exponentially flat potentials \eqref{eq:expU} and sufficiently large oscillation amplitudes, $W$ can be approximated by Eq.~\eqref{eq:W_expU} so that \eqref{eq:appr_peak} gives
\be\label{eq:mu_expU}
    \mu_k = (k/a) \sqrt{\frac{\sqrt{2}C}{\pi} \mth T - 1} + \mathcal{O}(k/a)^3\, .
\ee
In the special case of the potential \eqref{eq:U} with $n=2$ this expression with $C=1$ holds for any amplitude (see appendix \ref{app:n=2}).

For large $k$, $\tr G/2 \to \cos(kT/a)$ (see appendix \ref{sec:mu_asymp}), as can be observed from the lower panels in Fig.~\ref{fig:muk_U}. It implies that $\mu_k \to 0$ when $k\to \infty$, as expected. However, the way $\mu_k$ approaches 0 depends non-trivially on the behaviour of background oscillations. General arguments show that it must approach zero at least as $k^{-1}$~\cite{1968ZaMM...48R.138R}. However, our numerical estimates for the class of potentials \eqref{eq:expU} show that $|\tr G |-2$ is suppressed exponentially as $k$ grows. As a result, the instability bands get exponentially narrower and weaker when $k$ approaches infinity.

\subsection{Energy density of perturbations and backreaction} 
\label{sec:backreaction}

The energy density of perturbations can be estimated as the quantum average of the corresponding energy density operator (see appendix \ref{sec:mode_evol})
\bea\label{eq:pert_rho}
    \delta\rho(t) 
    = \int \frac{\td^3 k}{(2\pi)^3} \delta \rho_k(t) \, , 
    \qquad \mbox{where} \qquad
    \delta\rho_k 
    \equiv \frac{1}{2}|\delta\dot{\phi}_k|^2 + \frac{1}{2}\omega_k^2|\delta\phi_k|{}^2
\eea
is the energy density of a single mode. Assuming adiabaticity and neglecting expansion, the time averaged energy density of the perturbations grows as
\be\label{eq:drho_growth}
    \delta \dot{\rho}_k(t) = 2 \mu_k(\rho) \delta \rho_k(t) 
\ee
by Eq. \eqref{eq:muT_def}. As discussed above, mode growth depends on the potential and the oscillation amplitude $\phi_\amp$, or equivalently, the background energy density $\rho$.\footnote{Note that $\rho$ is reserved for the background energy density throughout the paper.}

We assume initial conditions corresponding to the vacuum state, which we take to be of the Bunch--Davies form, $\delta \rho_k^{\rm vac} = \delta \rho_k^{\rm BD} = k a^{-4}/2$.\footnote{We neglected the contribution from $U''$ which induces at most an order one correction. In the numerical approach, we set $a=1$ at the start of preheating. } As is well known, the energy density is divergent in the ultraviolet, so the integral \eqref{eq:pert_rho} must be regularized. In the following, we will use a sharp or an exponential ultraviolet cut-off. In particular, a sharp cut-off is implicit in our numerical approach below as the mode equations are solved for a finite set of momenta. Furthermore, we must renormalize the energy density of perturbations. We do this by demanding that it vanishes in the initial vacuum state. The physical, renormalized energy density $\delta\rho^{(R)}$ of a single mode then reads
\be\label{eq:rho_R_def}
    \delta \rho^{(R)}_k = \delta\rho_k - \delta\rho^{\rm vac}_{k} \, ,
\ee
where $\delta\rho^{\rm vac}_{k} = \delta\rho_{k}(0)$ is the initial regularized vacuum energy density for the mode. Essentially, Eq.~\eqref{eq:rho_R_def} renormalizes the cosmological constant to its observed value, which is negligible during preheating.

Above we considered mode growth neglecting the evolution of the background and the expansion of space. To remedy this, we impose energy conservation for the combined system of perturbations and background, so that, by Eq.~\eqref{eq:drho_growth}, energy is transferred from the background to the perturbations with the rate
\be\label{eq:energy_transfer_rate}
    s(t) \equiv \int \frac{\td^3 k}{(2\pi)^3} 2\mu_{k} \delta\rho_k(t) \, .
\ee
It matches $\delta\dot\rho^{(R)}$ when expansion is negligible. 

It is instructive to split the total growth rate into two terms
\be\label{eq:pert_rho_rate}
     s(t) = \bar \mu (\rho) \delta\rho^{(R)}(t) + j(\rho) \, ,
\ee
where we defined
\be\label{eq:murho_j}
    \bar \mu(\rho)
    \equiv \int \frac{\td^3 k}{(2\pi)^3} 2 \mu_k(\rho) \frac{\delta \rho^{(R)}_k(t)}{\delta\rho^{(R)}(t)} \, ,
    \qquad
    j(\rho)
    \equiv  \int \frac{\td^3 k}{(2\pi)^3} 2 \mu_k(\rho) \delta\rho^{\rm vac}_{k}\, .
\ee
In this formulation, the $j$ term acts as an explicit quantum source for the perturbations\footnote{The integral defining \eqref{eq:murho_j} converges due to $\mu_k(\rho)$ being exponentially damped at large $k$.}, originating from vacuum fluctuations, and dominates initially when the energy density of perturbations is negligible. Later, the effective growth rate $\bar \mu$ begins to drive the exponential growth of $\delta\rho^{(R)}$. Note that, on top of $\rho$, $\bar{\mu}(\rho)$ and $j(\rho)$ depend on the spectrum of perturbations. Therefore, one must still solve the evolution mode by mode. However, in the idealized limit of a constant background density, one expects that
\be\label{eq:mubar_to_mupk}
    \lim_{t \to \infty}\bar \mu = 2\mu_{\pk},
\ee
since the $k_{\pk}$-mode grows the fastest and will eventually dominate the energy density of perturbations. We will return to this limit in section~\ref{sec:approx_backreaction}.

On top of the energy transfer from the background to the fluctuations, both $\rho$ and $\delta\rho$ are diluted by Hubble friction; for $\rho$, the dilution follows from \eqref{eq:rho_friction_evolution}, while $\delta\rho$ dilutes in a $k$-dependent way. We can then describe the time-averaged evolution of all components of the scalar's energy density as 
\bea\label{eq:evol_rho+deltarho2}
    \dot{\rho} + 3 H W(\rho)/W'(\rho)   &= -s(t)  \, , 
    \\
    \delta\dot{\rho}_k + c_k H \delta \rho_k  &= 2\mu_k(\rho) \delta \rho_k \, ,
\eea
with expansion determined from the physical energy density by
$
    3H^2 = \rho + \delta \rho^{(R)}.
$
The factors $c_k$ capture the dilution of the $k$-modes. For massive modes, $c_k=3$, while for effectively massless high-$k$ modes, $c_k=4$; we expect $c_k$ to vary between these values. To capture the exact behaviour, $c_k$ would need to be solved from the full mode equation \eqref{eq:deltaphi_eq}. In the numerical approach below, we are content with adequately describing the leading modes, for which the mass term $U''$ is important, and thus we choose $c_k=3$ for all $k$.\footnote{Lattice studies in the parametric resonance regime show that the fragmented field behaves as radiation when $n>2$~\cite{Lozanov:2016hid,Lozanov:2017hjm} indicating that $c_k = 4$ would better capture the dilution after fragmentation.} For consistency, we take the vacuum modes $\delta\rho_k^{\rm vac}$ in \eqref{eq:rho_R_def} to dilute the same way, so that they correspond to a non-growing solution with $\mu_k=0$.

As long as the adiabatic approximation is valid, Eqs.~\eqref{eq:evol_rho+deltarho2} can be employed to study more extended time periods, during which Hubble friction can be relevant. However, as we will see below, in the parameter region allowing for tachyonic preheating, the feedback from perturbation growth is much stronger than Hubble friction, \ie, $\bar \mu \gg H$, and thus the latter plays an insignificant role.

All of the above relies on linear perturbation theory and will change when non-linearities become relevant. Interactions between the highly exited modes will eventually lead to the thermalization of the fragmented field. As this process redistributes the energy carried by the modes of the fragmented field, it does not strongly affect the total energy density in the fragmented field. However, as it can move modes in or out of instability bands, it could have a sizable effect on the fragmentation process itself. In particular, even if the power spectrum develops a dominant peak centred around $k_\pk$ as predicted by \eqref{eq:mubar_to_mupk}, it will eventually broaden due to non-linear effects. This process, including the generation of secondary peaks due to rescattering, has been observed in lattice studies~\cite{Lozanov:2016hid,Lozanov:2017hjm}.

\subsubsection{The dominant peak approximation}
\label{sec:approx_backreaction}

Time evolution simplifies if mode growth is dominated by a single peak in $\mu_k$, as is the case for strongly tachyonic preheating. Around the peak mode $k_\pk$, we can use the quadratic expansion
\be\label{eq:muk_quadratic}
    \mu_k \approx \mu_\pk - \frac{\Delta}{2a^2} (k - k_\pk)^2 \, 
\ee
to describe the first instability band and neglect mode growth in other instability bands.\footnote{The growth rate is positive as long as $|k - k_\pk|/a < \mu_\pk \sqrt{2/\Delta}$. Applying this approximation away from this range leads to unphysical damping of the modes. However, as long as $k_\pk$ does not evolve too fast, these far-off modes remain subdominant and can thus be neglected due to the exponential nature of perturbation growth. In this case, extending the approximation \eqref{eq:muk_quadratic} also to the far-off modes does not alter the estimate of the total energy density of perturbations.}

Consider first the simplified case where we neglect both expansion and backreaction. 
Then $\mu_k$ is time independent and
\be
     \delta\rho_k = \delta \rho_{k}^{\rm BD} e^{2\mu_k t} \, ,  
\ee
where $\delta\rho_{k}^{\rm BD}$ is the initial vacuum energy. The quadratic approximation \eqref{eq:muk_quadratic} thus produces a Gaussian peak in the perturbation spectrum with the width $(\Delta t)^{-1/2}$ that decreases in time. Relying on \eqref{eq:k0growth}, we can use $\mu_0 = 0$ to estimate $\Delta \approx 2\mu_\pk a^2/k_\pk^2$. In the limit $2\mu_\pk t \gg 1$, the total energy density of perturbations \eqref{eq:pert_rho} is then approximately\footnote{Note that in the limit of large perturbations, $\delta\rho_k \gg \delta\rho_k^\mathrm{BD}$ and thus $\delta\rho_k^{(R)} \approx \delta\rho_k$.}
\be\label{eq:rho_simple}
    \delta\rho^{(R)} 
    \approx \frac{(k_{\pk}/a)^4}{4\pi^{3/2}\sqrt{2\mu_\pk t}} e^{2\mu_\pk t} .
\ee
In this limit, $\mu_\pk$ clearly determines the growth rate as conjectured in \eqref{eq:mubar_to_mupk}. The time $\tau$ for the field to half-fragment, \ie, to reach $\delta \rho^{(R)}(\tau) \equiv U_0/2$ is
\be\label{eq:tau_simple}
    \tau \approx \frac{1}{\mu_\pk}\left[ \ln\left(\frac{4.0 \sqrt{U_0}}{(k_{\pk}/a)^2}\right) +\frac{1}{4} \ln \ln\left(\frac{4.0 \sqrt{U_0}}{(k_{\pk}/a)^2}\right) \right] \, .
\ee
Fixing $\mth$ \eqref{eq:m_value}, the initial amplitude \eqref{eq:phi_i} and the corresponding half-period of oscillations \eqref{eq:T_i} and using the large amplitude estimate \eqref{eq:appr_peak} for $\mu_\pk$ and $k_\pk$, we obtain that the field will half-fragment within
\be\label{eq:N_simple}
    N \approx \tau H \approx \mathcal{O}(1) \sqrt{\phi_0}
\ee
$e$-folds. The $\mathcal{O}(1)$ factor depends on $n$ and contains $\ln \phi_0$ corrections, \eg, for $n=2$ and $\phi_0 = 10^{-4}$ it is 1.1. These specifics are accounted below in Fig.~\ref{fig:Ngrid}, where this simple approximation is compared with more accurate estimates for half-fragmentation.

The number of half-oscillations before half-fragmentation is small. Following the procedure above, it is approximately
\be\label{eq:tau/T_simple}
    \tau/T \approx \frac{10}{-\log_{10}(\phi_0)}\, .
\ee
It follows that the field can become significantly fragmented already after the first oscillation when $\phi_0 \ll 10^{-5}$. We note, however, that Floquet theory gives the growth rate after a single half-oscillation and can thus be applied even if the number of half-oscillations is small (given the backreaction effects stay irrelevant).

To improve the estimate, let us include the possibility of a time dependent background. The approximation \eqref{eq:muk_quadratic} implies that the perturbation spectrum evolves as a Gaussian peak of the form
\bea\label{eq:delta_rho_gaussian}
    \delta\rho_k = \delta \rho_{k}^{\rm BD} \, e^{A - B(k - k_0)^2}, \,
\eea
and the ($H \to 0$) evolution of $\delta\rho$ is given by
\be\label{eq:gaussian_coefficient_eom}
    \dot{k}_0 = -\frac{\Delta}{a^2 B}(k_0 - k_\pk) \, , \quad
    \dot{B} = \frac{\Delta}{a^2} \, , \quad
    \dot{A} = 2\mu_\pk - \frac{\Delta}{a^2}(k_0 - k_\pk)^2 = 2\mu_{k_0} \, ,
\ee
where $k_\pk$, $\mu_\pk$ and $\Delta$ depend on the background energy density $\rho$ and thus on time. The time evolution drives the centre of the distribution $k_0$ towards the fastest-growing mode $k_\pk$. As in the time-independent case, the width $B^{-1/2}$ decreases monotonously---the fastest-growing modes outpace all others and the distribution gets narrower and narrower as time passes. The parameter $A$ captures the growth of the $k_0$ mode. Since the perturbations should start from the vacuum with $\delta\rho_k = \delta \rho_{k}^{\rm BD}$, we choose the initial conditions $A(0) = 0$, $k_0(0) = k_{\pk,i}$ (the initial fastest-growing mode), and $B(0)$ sufficiently small (the precise choice for $B(0)$ is discussed below).

In the limit $B \gg 1/k_0^2$, we can replace $0\to-\infty$ in the momentum integral \eqref{eq:pert_rho}, so the total perturbation energy density is
\be\label{eq:integrated_delta_rho}
    \delta\rho 
    \approx \frac{k_0}{4(\pi B)^{3/2}}\qty(\frac{3}{2} + B k_0^2)e^A  \, .
\ee
Since $B$ increases fast, the induced error is never large in practice. From Eq.~\eqref{eq:gaussian_coefficient_eom} we then obtain
\bea\label{eq:mu_rho}
     \delta\dot\rho 
     &\equiv s(t) 
     \approx \delta\rho \times 
\\ & \Bigg(2\mu_\pk - \frac{\Delta}{a^2}\qty[\frac{9 + 2Bk_0^2}{2B(3+2Bk_0^2)} + \frac{3+6Bk_0^2}{Bk_0(3+2Bk_0^2)}(k_0 - k_\pk) + (k_0 - k_\pk)^2]\Bigg) \, .
\eea
The first term in the brackets arises from the fastest-growing mode and the second term corrects for the finite (and changing) width of the distribution, and we identified the result with the source term $s(t)$ from \eqref{eq:energy_transfer_rate}.

With these results, we can numerically integrate the coupled evolution of the background and perturbations \eqref{eq:evol_rho+deltarho2} by using \eqref{eq:mu_rho} for $s(t)$, replacing the $k$-equations with \eqref{eq:gaussian_coefficient_eom}, and using $\delta\dot\rho + 3H\delta\rho = s(t)$ to include Hubble friction for the perturbations. We still use $\delta\rho^{(R)}=\delta\rho - \delta\rho^\mathrm{vac}$, where $\delta\rho^\mathrm{vac}=\delta\rho(0)$, to renormalize the perturbation energy density for the Friedmann equation (note that in this scheme, for a positive $B$, all the integrated quantities are finite). These equations must still be solved numerically, but we have reduced the number of equations from one for each $k$-mode to a total of five, with the cost of some accuracy.
 
We remark that, when we apply \eqref{eq:mu_rho} at earlier times when the perturbation spectrum is not narrow, we are effectively using the approximation \eqref{eq:muk_quadratic} in the region $|k - k_\pk| > 2\mu_\pk a^2/\Delta$ where $\mu_k<0$. If these modes would ever dominate, even $s(t)<0$ can be possible due to the unphysical damping of vacuum energy at high $k$. Thus, the quadratic approximation \eqref{eq:muk_quadratic} can only apply if these modes are never relevant during the time evolution of the system. In particular, the initial $B$ should be sufficiently large so the leading modes in the Gaussian peak \eqref{eq:delta_rho_gaussian} all reside inside the $\mu_k>0$ region; this also guarantees the validity of \eqref{eq:integrated_delta_rho} and \eqref{eq:mu_rho}. On the other hand, the initial vacuum state should have a vanishing $B$. As a compromise, we choose the initial $B$ to be the minimal value that still produces $s(0)\geq0$. The initial $\mu_k<0$ modes must also stay unimportant at later times; to achieve this, the background oscillation amplitude, and consequently $k_\pk$ and $\Delta$, should only change a little during the bulk of preheating. This approximation is then only a slight improvement over the simplified case with no background evolution considered at the beginning of this section but may be useful close to the parameter region where the simplified approximation starts to fail.

\subsection{Results}
\label{sec:results}

We now solve the evolution of the background and the perturbations using the tools discussed above to compute the duration of preheating and the ensuing perturbation spectrum. We perform the computation in multiple ways, taking varying amounts of computational resources. In all cases, we assume adiabatic background evolution so that the background's energy density $\rho$ decreases only a little within a single oscillation. In the order of increasing simplicity and decreasing accuracy, our approaches are:
\begin{itemize}
    \item \textbf{Grid}: We numerically integrate \eqref{eq:evol_rho+deltarho2} over a grid of $k$-values. To facilitate the computation, we first build the function $\mu_k(\rho)$ by numerically solving $\mu_k$ as described in section \ref{sec:mode} for several values of $\rho$ and interpolating in-between, for all $k$-values on the grid. We use a logarithmic grid with steps of $\Delta \ln k = 0.05$ and covering the range $k=10^{-15}\mth \dots 10^5 \mth$, chosen to capture all important modes that have a sizeable $\mu_k$ and may get excited during tachyonic preheating.\footnote{In practice, $\mu_k=0$ within the accuracy of the computation for all modes higher than the cut-off value on the grid. All higher $k$ modes stay in their vacuum state throughout the simulation, and the results are not sensitive to the cut-off. We tested the convergence of the results by repeating chosen computations with a doubled grid spacing, which produced only a negligible difference.}
    
    \item \textbf{Dominant peak approximation (DP):} We compute the evolution of perturbations with a time-dependent background as described in section~\ref{sec:approx_backreaction} by solving Eq.~\eqref{eq:mu_rho} with Hubble friction and backreaction included. In this setup, $\mu_k$ is given by the quadratic approximation \eqref{eq:muk_quadratic}. The coefficients $\mu_\pk(\rho)$, $k_\pk(\rho)$ and $\Delta(\rho)$ are obtained by fitting the first tachyonic peak. The contributions of the secondary peaks are neglected. As above, we build these functions by first finding $\mu_k(\rho)$ them for different values of $\rho$, \eg, as in Fig.~\ref{fig:peak}.
    
    \item \textbf{Simple approach:} We use the simplified version of the dominant peak approximation which relies on Eqs.~ (\ref{eq:rho_simple}-\ref{eq:N_simple}) and the fits \eqref{eq:appr_peak} to compute $N$ and the leading $k$. In this approximation, we assume a constant background energy density, that is, we neglect both backreaction and Hubble friction. Numerical integration is not needed in this approach.
    
\end{itemize} 
In the regime of tachyonic preheating, preheating finishes in much less than one $e$-fold (see Fig.~\ref{fig:Ngrid}). As a consequence, Hubble friction and redshift effects play a subleading role. Time evolution of $\rho$ is then mainly driven by the backreaction effects, as anticipated above. We neglect the redshift of the wavenumber $k$ in $\mu_k$ to simplify all the computations.

\begin{figure}
    \centering
    \includegraphics[scale=0.95]{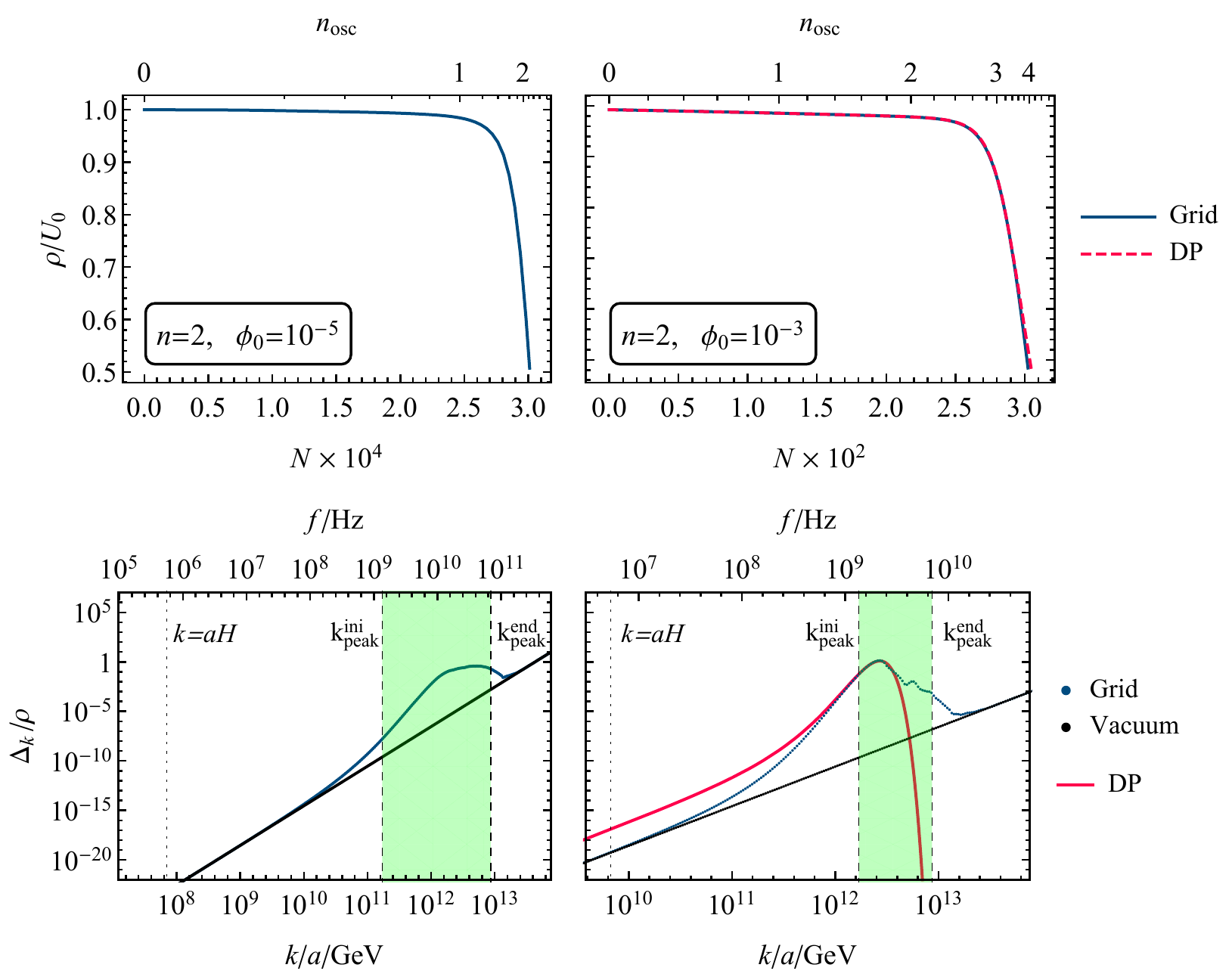}
    \caption{Evolution of the background energy density $\rho$ and the final energy spectrum of perturbations for $n=2$, for two example cases $\phi=10^{-5}$ (left) and $\phi=10^{-3}$ (right). For $\phi_0=10^{-3}$, the dominant peak approximation is also presented. The time evolution is shown in terms of the number of $e$-folds $N$ and the number of background half-oscillations $n_\mathrm{osc}$. Since the energy density of perturbations grows exponentially, the backreaction-driven time evolution is slow at first but accelerates towards the end. For the spectrum, the dotted line correspond to the Hubble horizon and the dashed lines to the value of $k_\pk$ initially and in the end. The corresponding frequencies today are also given. The spectrum is given in terms of $\Delta_k \equiv \frac{k^3}{2\pi^2}\delta\rho_k$, so that $\delta\rho = \int_0^\infty \td \ln k \, \Delta_k$.
    }
    \label{fig:rho_and_spectrum}
\end{figure}

Consider the initial and final configurations of the system. After fixing $n$ and setting $\mth$ from \eqref{eq:m_value}, the potential \eqref{eq:U} contains a single free parameter $\phi_0$. This sets the initial oscillation amplitude $\phi_{\amp,i}$ and the corresponding background energy density $\rho$ through Eq.~\eqref{eq:phi_i}. Starting from the vacuum state with $\delta\rho^{(R)}=0$, we follow the time evolution of the system until half-fragmentation $\delta\rho^{(R)}=\rho$, that is, until the perturbations have drained half of the background energy and preheating has essentially completed. Around this point, our linear analysis breaks down, and subsequent time evolution needs to be analysed by lattice methods. If Hubble friction is negligible, the final value of the background oscillation amplitude is $\phi_\amp\approx \phi_0$ (with some $n$-dependence), solved from $U(\phi_\amp) = U_0/2$. At this point, we end the simulation and record the number of $e$-folds $N$ and the perturbation spectrum. We scan over a range of $\phi_0$ values and repeat the analysis for $n=2$, $4$, and $6$.

Two representative examples with $\phi_0=10^{-3}$, $\phi_0=10^{-5}$ and $n=2$ are shown in Fig.~\ref{fig:rho_and_spectrum}. Although the duration in $e$-folds grows by two orders of magnitude when $\phi_0$ grows from $=10^{-5}$ to $10^{-3}$, the background energy density follows a roughly similar exponential decay curve in both cases. The lower panel of Fig.~\ref{fig:rho_and_spectrum} indicates that the structure of the instability bands can change considerably during preheating. The position of the fastest-growing mode $k_\pk$ changes by about an order of magnitude when $\phi_0=10^{-3}$ and by almost two orders of magnitude when $\phi_0=10^{-5}$.

In particular, when $\phi_0=10^{-3}$, the peak in the perturbation spectrum from the grid computation matches well the estimate from the dominant peak method, confirming its usefulness. Initially, $k_\pk$ is located at $k/a = 1.1 \times 10^{12}$ GeV, but as the background field amplitude decreases, the peak and growth shift to larger wavenumbers. However, this shift is small: at the end of the simulation, the maximum of the spectrum lies at $k/a = 3.0 \times 10^{12}$ GeV, and the spectrum maintains a sharp, peaked form. This justifies using the dominant peak approximation.

\begin{figure}[p]
    \centering
    \hspace{-7mm}
    \includegraphics[scale=0.85]{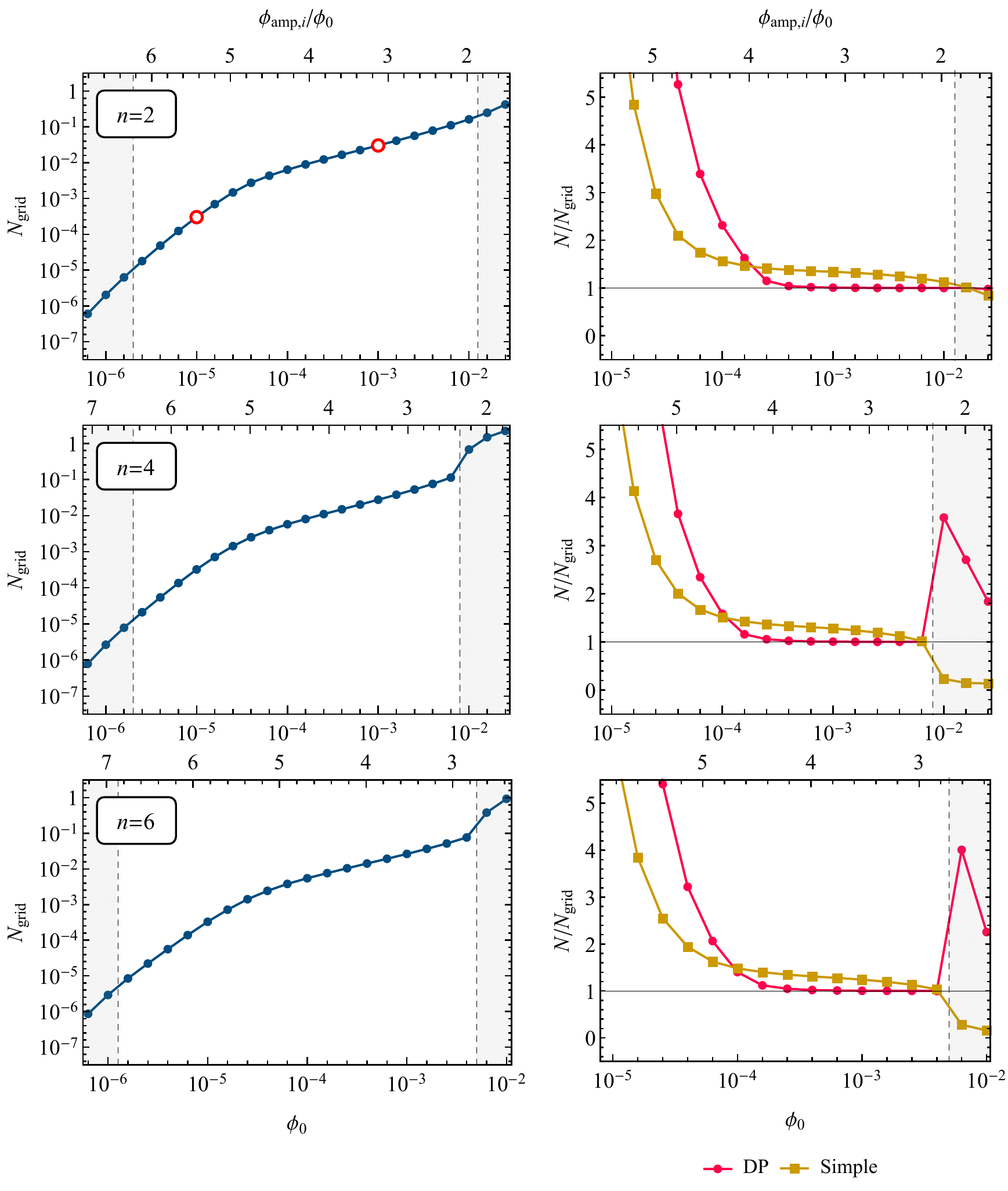}
    \caption{Duration of preheating in $e$-folds $N$, from the beginning of the oscillations to the moment at which the energy density of perturbations overtakes the background, as a function of the model variable $\phi_0$, for different $n$. The corresponding initial oscillation amplitude $\phi_{\amp,i}$ is also shown. \emph{Left:} Estimate from a numerical grid. The benchmark points used in Fig.~\ref{fig:rho_and_spectrum} are marked in red. \emph{Right:} Comparison of the grid result to the dominant peak and simple approximations in their region of validity $\phi_0 \gtrsim 10^{-4}$. In the shaded regions, the results become unreliable due to breakdown of adiabaticity for $\phi_0 \lesssim 10^{-6}$ and end of tachyoncity for $\phi_0 \gtrsim 10^{-2}$.}
    \label{fig:Ngrid}
\end{figure}

\begin{figure}[p]
    \centering
    \hspace{-7mm}
    \includegraphics[scale=0.85]{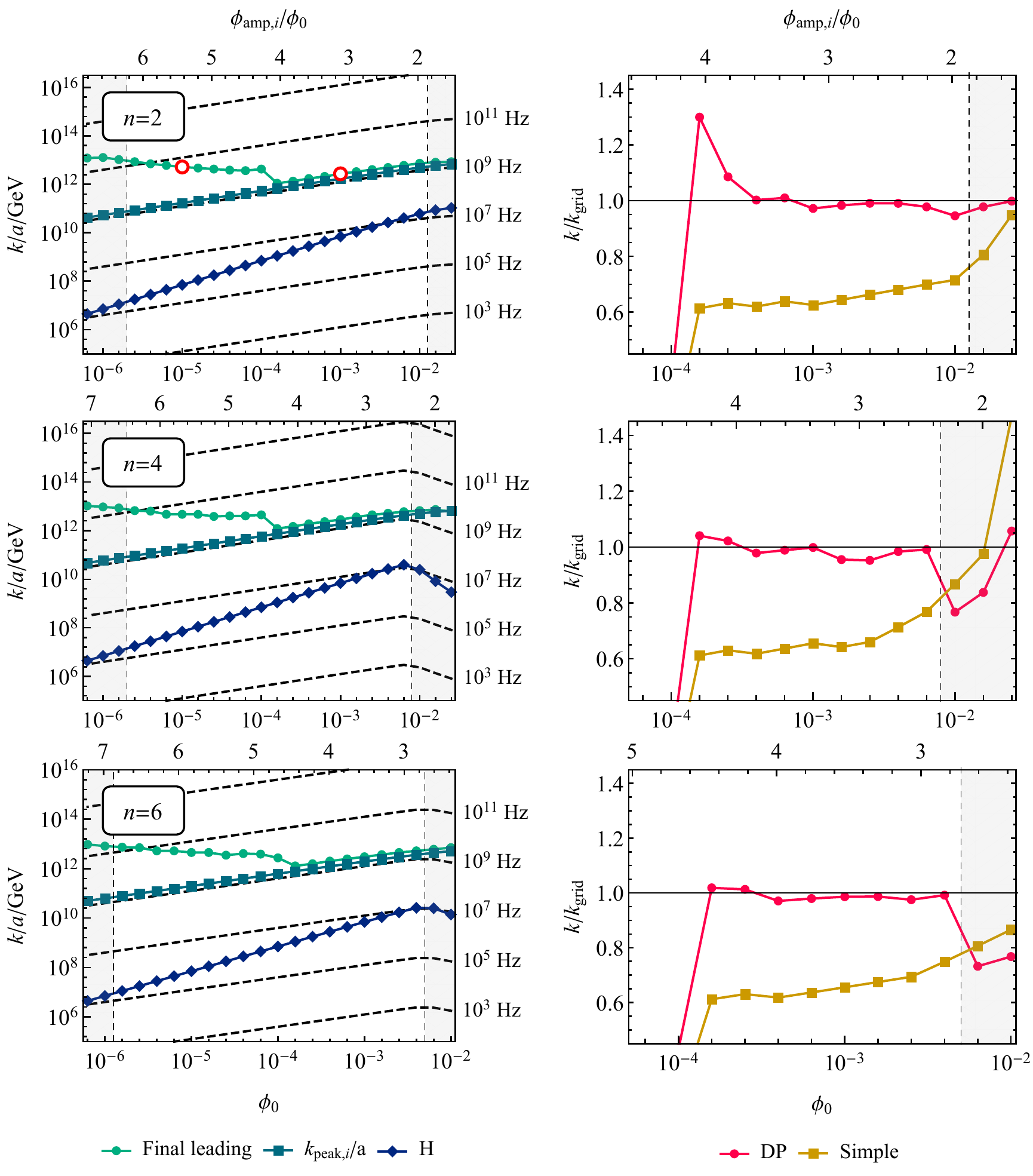}
    \caption{Different energy scales as functions of $\phi_0$. The initial oscillation amplitude $\phi_{\amp,i}$ is also shown. \emph{Left:} The results using a numerical grid showing the leading mode $k$ contributing most to the energy density at the end of preheating, the initial fastest-growing mode $k_{\pk,i}$, and the Hubble scale. The dashed lines indicate the present frequency of potential GWs generated during preheating, assuming a quick transition to radiation domination and a standard expansion history after that. The benchmark points used in Fig.~\ref{fig:rho_and_spectrum} are marked in red.
    \emph{Right:} Comparison of the final peak frequency on the grid and in the dominant peak and simple approximations.}
    \label{fig:kgrid}
\end{figure}

\afterpage{\FloatBarrier}

For $\phi_0=10^{-5}$, the peak in the power spectrum is much flatter, so the dominant peak and the grid-based results differ significantly, which is why Fig.~\ref{fig:rho_and_spectrum} only presents the more accurate grid-based solution. The difference is due to the larger range of $k$-values amplified during preheating. Initially, the fastest-growing mode is $k/a=1.1\times10^{11}$ GeV, while the final peak in the spectrum sits at $k/a=4.5\times10^{12}$ GeV, and the spectrum is flatter than in the previous case. One reason for this is that, for larger $\phi_{\amp,i}/\phi_0$, the secondary instability bands have a relatively larger $\mu_k$, as can be seen in Fig.~\ref{fig:muk_U}, and start to contribute to the growth of the energy density of perturbations. In addition, the oscillation amplitude changes more, so the tachyonic resonance band scans over a wider range of modes. It follows that the modes at the final peak are initially amplified by the weaker parametric resonance and then by the stronger tachyonic instability at the end of preheating as the first peak shifts towards smaller scales. The dominant peak approximation does not capture the evolution of modes in the secondary instability bands and thus severely underestimates preheating efficiency.

The simplified approximations fail when $\phi_0 \lesssim 10^{-4}$. We find an upper bound $\phi_0 \lesssim 10^{-2}$ for preheating to be driven by the tachyonic instability, which is consistent with earlier lattice computations ~\cite{Lozanov:2016hid, Lozanov:2017hjm}. When $\phi_0 \gtrsim 10^{-2}$, the initial amplitude is small, so the background exits the tachyonic region fast and reheating must complete through other channels (right shaded region in Figs.~\ref{fig:Ngrid} and~\ref{fig:kgrid}). At the other end of the scale is the lower bound $\phi_0 \gtrsim 10^{-6}$ due to adiabaticity -- in this case, preheating completes within less than a half-oscillation~\footnote{We estimate the number of oscillations as $n_\mathrm{osc} = \int \td t/T(\rho)$ in the presence of an evolving background density, \ie, when $\dot\rho \neq 0$.} (left shaded region in Figs.~\ref{fig:Ngrid} and~\ref{fig:kgrid}). In fact, adiabaticity begins to be violated at somewhat larger values $\phi_0 \lesssim 10^{-5}$, in the sense that quantities like $T \dot{\phi}_\amp/\phi_\amp$ are order one, so the amplitude changes considerably during a half-oscillation. For $10^{-5} \lesssim \phi_0 \lesssim 10^{-2}$, this quantity stays below unity almost until the end of the simulation, although the total number of half-oscillations is never high; about $5$ for $\phi_0 \sim 10^{-2}$. These order-of-magnitude estimates apply for all our $n$-values.

Figs.~\ref{fig:Ngrid} and~\ref{fig:kgrid} show the duration of preheating and the position of the peak of the final spectrum for varying $\phi_0$, comparing the grid-based estimate, the dominant peak approximation and the simple estimate. The value of $N$ varies between $10^{-5}$ and $10^{-1}$ in the eligible region and goes down for smaller $\phi_0$. For the leading final $k$, this is also true for $\phi_0 \gtrsim 10^{-4}$, but for lower $\phi_0$, the peak value starts to grow again due to the secondary, higher-$k$ peaks becoming more prominent. All relevant $k$-values are well within the Hubble radius, more so for lower $\phi_0$. These results agree with the earlier findings for $n=4$~\cite{Rubio:2019ypq} and $n=2$~\cite{Karam:2021sno}.

Both the dominant peak approximation and its simplified version predict values of $N$ and the final leading $k$ that are of the correct order of magnitude for $10^{-4} \lesssim \phi_0 \lesssim 10^{-2}$, where the initial and final leading $k$ are close to each other. As expected, the dominant peak approximation performs better, being within a few percent of the grid result in most of this region, but even the simple approach produces results within a factor of two of the grid value for $k$ and $N$. The simple approach slightly underestimates the leading $k$ and the preheating efficiency. Beyond their region of validity, the approximations start to deviate from the grid, severely underestimating $k$ and overestimating $N$.

\subsection{Discussion}

Consider the energy scales for preheating implied by the approximations given above. Throughout the discussion we will use $\mth = 1.2 \times 10^{13} \, \GeV$ fixed by the CMB measurement \eqref{eq:m_value}. Eqs. \eqref{eq:appr_peak} and \eqref{eq:T} tell that the position of the dominant peak at the moment of preheating is
\be
    k_\pk/a_{\rm preh} \approx \sqrt{\phi_0} \, 9 \times 10^{13}\, \GeV \, ,
\ee
in the region $10^{-4} \lesssim \phi_0 \lesssim 10^{-2}$ in which the dominant peak approximation works well. This expression has a mild dependence on $n$. The above value was computed for $n=2$.  When $10^{-6} \lesssim \phi_0 \lesssim 10^{-4}$, the perturbation spectrum will peak at higher frequencies as can be seen from Fig.~\ref{fig:kgrid}. Assuming instantaneous thermalization, we can approximate $\rho_R \approx U_0$, so the temperature after thermalization
\be
    T_{\rm preh} \approx \sqrt{\phi_0} \, 2 \times 10^{15}\, \GeV 
\ee
exceeds the peak momentum by more than an order of magnitude, suggesting that thermal effects will shift the initial spectrum towards higher modes.

Since tachyonic preheating is such a violent process, one may expect strong production of secondary GWs. As Figs.~\ref{fig:kgrid} and \ref{fig:rho_and_spectrum} show, a potential GW signal is created in the 1-10 GHz range when $\phi_0=10^{-4}-10^{-2}$. This range is almost independent of $\phi_0$ or $n$. This frequency can be estimated by redshifting the peak of the spectrum by a factor $a_{\rm now}/a_{\rm preh} = (g_{*s,\rm preh}/g_{*s,\rm now})^{1/3}T_{\rm preh}/T_{\rm now}$, where $T_{\rm now} = 2.7$ K is the present photon temperature~\cite{Fixsen:2009ug}. This gives
\be
    f_{\rm GW} 
    \approx k_\pk/a_{\rm now}
    \approx 5 \,\GHz
\ee
and the dependence on $\phi_0$ has dropped out. These numbers are compatible with the lattice studies~\cite{Lozanov:2019ylm, Bhoonah:2020oov} performed for similar models. Such frequencies are not observable with near-future gravitational-wave interferometers~\cite{Somiya:2011np,TheVirgo:2014hva,Martynov:2016fzi}, but may be probed through the effects on big bang nucleosynthesis~\cite{Pagano:2015hma, Ejlli:2019bqj,Domcke:2020yzq} or 21 cm measurements~\cite{Domcke:2020yzq,Ringwald:2020ist}.

Finally, let us take a closer look at the applicability of the linear theory for small $\phi_0$. There are several theoretical arguments that combined with our numerical results, point towards complications when $\phi_0 \lesssim 10^{-6}$:
\begin{itemize}
    \item To estimate whether perturbation theory may be applied, consider the effective quartic coupling $\lambda \sim U^{(4)}(\phi) \sim U_0/\phi_0^4 = \mth^2/\phi_0^2$. Since $\mth \approx 5 \times 10^{-6}$ is fixed by \eqref{eq:m_value}, the quadratic interaction $\lambda$ exceeds unity when $\phi_0 \lesssim 10^{-6}$. At this point, perturbativity is violated, and non-linear contributions should not be neglected even at the beginning of preheating when the energy density of fluctuations is negligible.
    
    \item Due to the shape of the potential \eqref{eq:U}, the model is non-renormalizable even when gravitational interactions are neglected. However, quantum corrections can be treated in an effective field theoretic framework, so that all relevant energy scales must stay below the cut-off scale of the effective field theory, which for the potential \eqref{eq:U} is approximately $\Lambda_\mathrm{cut} = \phi_0$. From Fig.~\eqref{fig:kgrid}, we see that the peak of the spectrum does not follow $k_{\pk,i}$ when $\phi_0 \lesssim 10^{-4}$. In fact, the peak of the perturbation spectrum scales roughly as $k/a\sim \mth$ rather than $k/a\sim \mth \sqrt{\phi_0}$ as was estimated in Eq.~\eqref{eq:appr_peak}. Our numerical estimates show that the position  $k$ of the peak of the spectrum rise above the cut-off when $\phi_0 \lesssim 10^{-6}$. The same unitarity bound, $U_0 \gtrsim \Lambda_\mathrm{cut}^4$, was derived in Ref.~\cite{Ema:2021xhq}, although not in the context of tachyonic preheating.
    
    \item When the (non-renormalized) vacuum energy density of modes that can contribute significantly to perturbations is larger than the background energy density, then only a mild amplification of the vacuum fluctuations can result in the total fragmentation of the field. An analogous argument was used in Ref.~\cite{Lozanov:2019ylm}, to derive the upper bound $4\times10^{10}$ Hz for the frequency of GWs from preheating. 
    
    Technically this means that $jT \gtrsim \rho$, so that the quantum source term $j$ defined in Eq.~\eqref{eq:murho_j} dominates. This is possible due to the contribution of higher instability bands.\footnote{Note that, due to Eq.~\eqref{eq:appr_peak}, the (non-renormalized) vacuum energy density around $k_{\pk,i}$ is always much smaller than the initial background energy density, $\delta\rho^{\rm vac}_{k_\pk} \sim k_\pk^4 \sim U_0 \mth^2 \ll U_0$.} As the condition implies significant energy transfer during a half-oscillation, it is essentially equivalent to the breakdown of adiabaticity. Indeed, neglecting backreaction and computing the evolution of linear perturbations on the grid from the mode equation \eqref{eq:deltaphi_eq}, $\delta\rho^{(R)}$ exceeds $\rho$ by the first zero-crossing of $\phi$ when $\phi_0 \lesssim 10^{-6}$, implying, at the very least, an exit from the oscillatory regime. We remark that the dominant peak estimate \eqref{eq:tau/T_simple} for the number of oscillations would severely underestimate the speed of fragmentation in this case.
    
\end{itemize}
Due to these complications, the parameter region $\phi_0 \lesssim 10^{-6}$ seems to behave in a fundamentally non-linear way and thus, a more involved analysis is indeed needed. The fact that only a mild amplification, or equivalently, the production of a few quanta leads to fragmentation can also be interpreted as the onset of a dominantly quantum regime. The implication of the latter is that the $\phi_0 \lesssim 10^{-6}$ region cannot be probed by classical lattice studies.

\section{Conclusions}
\label{sec:concl}

We have studied preheating in inflationary scenarios in which the inflaton possesses an exponentially flat plateau. We focused on the parameter region in which the fragmentation of the inflaton is mainly due to a strong tachyonic instability. Although perturbation growth was studied at the linear level, the usual linear approach was extended by including backreaction effects between the coherent oscillating background and the growing perturbations.

Tachyonic preheating in plateau inflation turns out to be a rapid process, lasting only a fraction of an $e$-fold. This permitted us to treat cosmic expansion adiabatically. Thus, at the leading order, the quantitative characteristics of preheating can be determined in a Minkowski background. We analytically derived characteristics of inflation, the oscillating background and linear mode growth, which are universal to exponentially flat potentials. In particular, the growth of individual modes lends itself to Floquet theoretic treatment, and using this method, we numerically computed the wavenumber and the growth rate of the fastest growing mode. In the tachyonic regime, these quantities depend only weakly on the exact shape of the exponentially flat potential.

During tachyonic preheating, the energy density of the oscillating background gets damped mainly because its energy is transferred to the fragmented component of the inflaton. The damping due to Hubble friction is always subdominant. As the background's oscillations proceed with ever decreasing amplitude, the structure of the instability bands, which determines the growth of each individual mode, changes. To account for the interplay between the evolving background and mode growth, we constructed a set of coupled continuity equations that appropriately account for the energy transfer between these components in the adiabatic limit. 

Mode growth in tachyonic preheating is characterised by a dominant instability band at the lowest wavenumbers. We showed that further simplification is possible when considering mode growth only from modes surrounding the fastest growing mode. Within this approximation, it is possible to construct analytic order of magnitude estimates that relate the global characteristics of preheating, \eg, its duration in $e$-folds, to the parameters of the model. Using more accurate numerical estimates, we confirmed that this approximation works well as long as adiabaticity is not violated.

In particular, we considered potentials of the form
$$
    U(\phi) = U_0 \tanh^{n}(\phi/\phi_0) \,
$$
with even $n$, as representative cases of the general class of exponentially flat potentials. For such potentials, CMB measurements fix $\mth \equiv \sqrt{U_0}/\phi_0 = 5\times10^{-6}$, the half-period of background oscillations is $T \approx \mth^{-1} \phi_0^{-1/2} \ll H^{-1}$, and the fastest growing mode as well as its growth rate are $k_\pk \approx \mu_\pk \approx \pi/T$. Since $\mu_\pk\sim T^{-1}$, preheating is expected to complete in a few oscillations. All of these expressions are nearly independently of $n$. The qualitative features are completely controlled by the parameter $\phi_0$. In particular, we find that
\begin{itemize}
    \item When $\phi_0 \gtrsim 10^{-2}$, then mode growth is driven by parametric resonance and the tachyonic instability is subdominant. 
    
    \item When $10^{-6} \lesssim \phi_0 \lesssim 10^{-2}$, preheating is dominated by the tachyonic instability, and the background evolves adiabatically even if backreaction effects are taken into account. In this regime, preheating completes in approximately $N \approx \sqrt{\phi_0} \lesssim 0.1$ $e$-folds. We find that the analytic estimates work best in the parameter region $10^{-4} \lesssim \phi_0 \lesssim 10^{-2}$, while, when $10^{-6} \lesssim \phi_0 \lesssim 10^{-4}$, the adiabaticity begins to fail. The latter leads to deviations between the simplified estimates considering only the dominant peak in the mode growth spectrum and the more accurate numerical estimates accounting for the exact growth spectrum.
    
    \item When $\phi_0 \lesssim 10^{-6}$, the background does not evolve adiabatically, and the field can fragment significantly before completing a single half-oscillation. Although the tachyonic instability would dominate preheating after inflation, the damping of the background would rapidly activate the parametric regime. In all, since the linear approach becomes meaningless, $\phi_0 \lesssim 10^{-6}$ is the region of non-linear preheating in which quantum effects become relevant. As a related point, we estimate that inflaton's self-interactions are strong enough to violate unitarity in this regime.
    
\end{itemize}

Our description inevitably fails near the end of preheating, when non-linear effects become important, and for $\phi_0 \lesssim 10^{-6}$ due to the loss of adiabaticity. Rapid preheating is still expected, but the details have to be resolved by a more comprehensive analysis. For $\phi_0 \gtrsim 10^{-6}$, this can be done with a more resource-intensive classical lattice computation, like in~\cite{Lozanov:2016hid, Lozanov:2017hjm, Krajewski:2018moi, Lozanov:2019ylm, Bhoonah:2020oov}. In the future, our results can inform such studies regarding the relevant time and energy scales. For $\phi_0 \lesssim 10^{-6}$, classical lattice simulations are not applicable, since non-linear effects come into play already in the quantum regime, after the production of only a few quanta. New techniques are needed to resolve the dynamics in this case.

Finally, preheating's violent fragmentation process is expected to produce high-frequency GWs in the 1-10 GHz range. This frequency range is roughly determined by the fastest growing mode $k_\pk \approx \pi/T$. Although we do not address the GW spectrum and energy density in this paper, looking for such signals can provide valuable information about the very first moments in the early universe.

\acknowledgments

This work was supported by the Estonian Research Council grants PRG803, PRG1055, MOBTP135, MOBJD381 and MOBTT5 and by the EU through the European Regional Development Fund CoE program TK133 ``The Dark Side of the Universe."

\clearpage
\appendix
\section{Treatment of linear perturbations}
\label{sec:mode_evol}

We treat the linear inflaton perturbations quantum mechanically, given by the field operator
\be\label{eq:delta_phi_operator}
    \delta\hat{\phi}(x) = \int \frac{\td k^3}{(2\pi)^{3/2}} \qty(\delta\phi_k \hat{a}_{\vec{k}} e^{i\vec{k} \cdot \vec{x}} + \delta\phi^*_k \hat{a}^\dagger_{\vec{k}} e^{-i\vec{k} \cdot \vec{x}}) \, , \quad \comm{\hat{a}_{\vec{k}}}{\hat{a}^\dagger_{\vec{p}}} = \delta^3(\vec{k}-\vec{p}) \, , \quad \hat{a}_{\vec{k}}\ket{0} = 0 \, .
\ee
The ladder operators $\hat{a}_{\vec{k}}$ define the vacuum sate $\ket{0}$ related to the mode functions $\delta\phi_k$. Time evolution is contained in the mode functions $\delta\phi_k$, which follow the equations of motion \eqref{eq:deltaphi_eq},
\be\label{eq:dphik_eom_appendix}
    \delta\ddot{\phi}_k + 3H\delta\dot{\phi}_k + \omega_k^2 \delta\phi_k = 0 \, , \qquad \omega_k^2 \equiv \frac{k^2}{a^2} + U''(\phi) \, .
\ee
and we use the Bunch--Davies initial conditions
\be\label{eq:bunch_davies}
    \delta\phi_k = \frac{1}{\sqrt{2k}a} \, , \qquad \delta\dot{\phi}_k = -i\frac{k}{a}\delta\phi_k \, ,
\ee
corresponding to the Minkowski-like adiabatic vacuum state. The expectation value of the energy density of perturbations in the vacuum state is now
\bea\label{eq:pert_rho_app}
    \delta\rho &= \frac{1}{2}\expval{\delta\dot{\hat{\phi}}{}^\dagger
    \delta\dot{\hat{\phi}}
    +
    \frac{1}{a^2}
    \nabla_i \delta\hat{\phi}{}^\dagger
    \nabla_i \delta\hat{\phi}
    +
    U''(\phi)
    \delta\hat{\phi}{}^\dagger
    \delta\hat{\phi}
    } \\
    &=
    \int \frac{\td^3 k}{(2\pi)^3}\qty(\frac{1}{2}|\delta\dot{\phi}_k|^2 + \frac{1}{2}\omega_k^2|\delta\phi_k|{}^2) \, ,
\eea
which we use in \eqref{eq:pert_rho}.

Note that we have only included inflaton perturbations and neglected metric scalar perturbations, even though these two are coupled at linear order. This approximately corresponds to working in the spatially flat gauge, where the effect of metric perturbations is minimized. In this gauge, the remaining metric fluctuations play a subleading role for the leading sub-Hubble modes in the tachyonic regime~\cite{Rubio:2019ypq, Karam:2021sno}, and we neglect them for simplicity.

In \eqref{eq:dphik_eom_appendix}, the function $\omega_k^2$ depends on the background, which (disregarding backreaction) follows the equations
\be\label{eq:phi_eom_appendix}
    \ddot{\phi} + 3H\dot{\phi} + U_{,\phi} = 0 \, , \qquad
    3H^2 = \frac{1}{2}\dot{\phi}^2 + U \, , \qquad U = U_0\tanh^n \frac{\phi}{\phi_0} \,
\ee
introduced in section \ref{sec:classical}, where we specialized to our model with the potential \eqref{eq:U}. To better understand the dependence of the dynamics on the input parameters $\phi_0$ and $U_0$, it is useful to introduce the rescaled variables
\be\label{eq:rescaled_vars}
    \tphi \equiv \frac{\phi}{\phi_0}\, , \quad \tlt \equiv t \, \mth \, , \quad E \equiv \frac{H}{\sqrt{U_0}} \, , \quad \tU \equiv \frac{U}{U_0} \, , \quad
    \tk \equiv \frac{k}{\mth} \, ,
\ee
where $\mth \equiv \sqrt{U_0}/\phi_0$ from \eqref{eq:mth}. In terms of these, the equations read
\begin{gather}
    \label{eq:bg_eqs_rescaled}
    \tphi'' + 3E\tilde{\phi}'\phi_0 + \tilde{U}_{,\tphi} = 0 \, , \qquad
    3E^2 = \frac{1}{2}\tphi'{}^2 + \tU \, , \\
    \label{eq:pert_eq_rescaled}
    \delta\phi''_\tk + 3E\delta\phi'_\tk\phi_0 + \qty(\frac{\tk^2}{a^2} + \tU_{,\tphi\tphi}) \delta\phi_\tk = 0 \, ,
\end{gather}
where a prime denotes a derivative w.r.t. the rescaled time $\tlt$. The potential height $U_0$ cancels out; it can be used to correctly fix the amplitude $A_s$ of CMB, but does not affect the dynamics. The parameter $\phi_0$ also only features in the friction terms, making them small, and does not enter in the $H\to0$ limit, \eg, when computing the growth rate $\mu_k$ \eqref{def:mu_k} in section~\ref{sec:mode}. Therefore, as $\phi$ can be measured in units of $\phi_0$ and energy and time scales are measured in units of $\mth$ (or a derived quantity like the oscillation time $T$), the explicit dependence on $\phi_0$, $U_0$, and $\mth$ can be removed by an appropriate rescaling of dimensionful quantities. In particular, combinations like $\mu_k T$ can be computed only by fixing a single background parameter, \eg, $\phi_\amp/\phi_0$.

\section{Asymptotics for \texorpdfstring{$H \to 0$}{H=0} mode growth}
\label{sec:mu_asymp}

In the following we will show that the growth of small $k$ modes for symmetric potentials is determined by\footnote{As expansion is irrelevant for the results of this section, we set $a=1$.}
\be\label{eq:k0growth_0th+1st_sym}
    \frac{1}{2} \Tr G = -1 - \frac{k^2}{2} W \partial^2_\rho W + \mathcal{O}(k^4) 
    \quad \Rightarrow \quad
    \mu_k = k \Re\frac{\sqrt{W \partial^2_\rho W}}{\partial_\rho W} + \mathcal{O}(k^2)
\ee
where $W$ is given by \eqref{def:W}.  For asymmetric potentials one must replace $\Tr G \to -\Tr G$ and $W$ by the corresponding integral over the full period. 

The large $k$ behaviour follows from treating the background adiabatically, which gives
\be
    \frac{1}{2}\Tr G = \cos\left(\int^{T}_0 \sqrt{k^2 + U''} + \mathcal{O}(k^{-2})\right)
    \quad \Rightarrow \quad
    \lim_{k\to \infty} \mu_k = 0 
\ee
with the damping of $\mu_k$ depending non-trivially on the shape of the potential.

\subsection*{The $k=0$ mode}

The equation of motion \eqref{eq:deltaphi_eq_noH} for the $k=0$ mode is
\be\label{eq:deltaphi_eq_noH_0}
    \delta\ddot{\phi}_0 - \frac{\dddot\phi}{\dot\phi} \delta \phi_0 = 0 \, ,
\ee
where we used that $U'' = -\dddot\phi/\dot\phi$ in a Minkowski background. It follows that the system has two independent solutions
\be\label{eq:u12,k=0}
    u_1 = \dot\phi, \qquad
    u_2 = \dot\phi \int \frac{\td t}{\dot\phi^2} \, ,
\ee
where the second solution follows from the fact that the time derivative of the Wronskian $\det w = u_1 \dot u_2 - u_2 \dot u_1$ vanishes. This set of solution satisfies $\det w = 1$. Hill's discriminant, \ie, the trace of the monodromy matrix \eqref{eq:G} is then
\bea\label{eq:k0growth_0th}
    \frac{1}{2}\Tr G^{(0)} 
&   = \frac{1}{2} \left( u_1(0) \dot{u}_2(T) - \dot{u}_1(0)u_2(T) + u_1(T) \dot{u}_2(0) - \dot{u}_1(T) u_2(0) \right)\\
&   = \pm \frac{1}{2}\left(u_1(T) \dot{u}_2(T) - \dot{u}_1(T)u_2(T) + u_1(0) \dot{u}_2(0) - \dot{u}_1(0) u_2(0) \right)
    = \pm 1\, ,
\eea
where on the second line we used $\det w = 1$ and $u_1(t+T) = \pm u_1(t+T)$, depending whether $T$ is the period or half-period of $\phi$ (depending whether the potential $U$ is asymmetric or symmetric, respectively). In both cases, this implies that the ground state mode does not grow,
\be
    \mu_{k=0} = 0 \, .
\ee
We remark that \eqref{eq:k0growth_0th} holds whenever the mode equation has a periodic solution -- the periodicity of the second solution $u_2$ is not necessary. This is also clear from $\det G = 1$, which implies that if one solution is oscillating, then the other solution must have a vanishing growth rate. Despite that, $u_2$ is not bounded and grows linearly. To see this we need to take a closer look at the singularity in its defining integral \eqref{eq:u12,k=0} at the turning point, \ie, when $\dot\phi = 0$. 

In the following, we will focus on symmetric potentials with $T$ the half-period, so that $\dot\phi(t+T) = -\dot\phi(t)$. Let us pick $t=0$ at the bottom of the potential, so that $\ddot\phi(0) = - U'(\phi_\amp) = 0$. The turning point $\phi_\amp$ is reached at $t = T/2$, so $\dot\phi(T/2) = 0$, and after that, the background field rolls down the potential with $\dot\phi(t) = -\dot\phi(T-t)$. In order to avoid integrating over the singularity, we define
\be\label{eq:u2_piecewise}
    u_2(t) =
\left\{\begin{array}{lc}
    \tilde u_2(t)  &, \quad            0 \leq t \leq T/2            \\
    \tilde u_2(T-t) + \beta \dot\phi(t) &, \quad   T/2 \leq t \leq T   
\end{array}\right.
\ee
with $\beta$ a constant and
\be
    \tilde u_2(t) \equiv \dot\phi(t) \int^t_{0} \frac{\td t'}{\dot\phi(t')^2}
\ee
defined in the range $0 \leq t \leq T/2$. With this choice of $u_2$, we have that
\be
    w(0) =
\begin{pmatrix}
    \dot\phi(0) & 0 \\
    0 & \dot\phi(0)^{-1}
\end{pmatrix} \, ,
\qquad
    w(T) =
\begin{pmatrix}
    \dot\phi(T) & \dot\phi(T) \beta \\
    0 & \dot\phi(T)^{-1}
\end{pmatrix}
\ee
and therefore, with $\dot\phi(T)=-\dot\phi(0)$, the monodromy matrix is given by
\be
    G_{k=0} \equiv w(0)^{-1}w(T) =
\begin{pmatrix}
    - 1 & -\beta \\
    0 & -1
\end{pmatrix}
\ee
indicating a linear growth of a $k=0$ mode when $\beta \neq 0$, \ie, after $n$ half-oscillations $(G^n)_{12} = (-1)^{n} n \beta$.

Our remaining task is to find the constant $\beta$. As long as $u_2$ is continuous and differentiable, \eqref{eq:u2_piecewise} will give a solution of \eqref{eq:deltaphi_eq_noH_0} satisfying $u_1 \dot u_2 - u_2 \dot u_1 = 1$. The only instance that needs to be checked separately is at $t=T/2$. First, at $t=T/2$, $\tilde u_2(T/2) = -1/\ddot\phi(t) = 1/U'(\phi_\amp)$ and $\dot{\phi}=0$ and thus $u_2$ is continuous as long as $U'(\phi_\amp) \neq 0$. The derivative $\dot u_2$ at $t=T/2$ is $\dot{\tilde u}_2(T/2)$ when $t \to T/2^{-}$ and $-\dot{\tilde u}_2(T/2) + \beta \ddot\phi(T/2)$ when $t \to T/2^{+}$. Thus, differentiability demands that
\bea
    \beta 
&   = \frac{2\dot{\tilde u}_2(T/2)}{\ddot\phi(T/2)}
    = \lim_{t\to T/2} \frac{2}{\ddot\phi(T/2)} \left[ \ddot\phi(t) \int^t_{0} \frac{\td t'}{\dot\phi(t')^2} + \frac{1}{\dot\phi(t)} \right] = - \partial_\rho T = - \partial_\rho^2 W\, .
\eea
The last two identities follow from \eqref{eq:T_and_P} and
\bea
    \partial_\rho T 
&   = 2\partial_\rho \int^{\phi_\amp}_{0} \frac{\td \phi'}{(2(\rho - U(\phi'))^{1/2}} \\
&   = \lim_{\phi \to \phi_\amp} 2 \left[ -\int^{\phi}_{0} \frac{\td \phi'}{(2(\rho - U(\phi))^{3/2}} + \frac{ \partial_\rho \phi_\amp }{(2(\rho - U(\phi))^{1/2}} \right] \\
&   = \lim_{t \to T/2} 2 \left[ -\int^{t}_{0} \frac{\td t'}{\dot\phi(t')^2} - \frac{1}{\dot\phi(t)\ddot\phi(t)} \right]\, .
\eea

\subsection*{Small $k$ modes}

To find the growth rates for small $k$, we look for perturbative solutions to the mode equation \eqref{eq:deltaphi_eq_noH} and expand
\be
    \delta \phi_k = \sum_i (-k^2)^{n} u^{(n)}
\ee
with each subsequent order determined iteratively by
\be
    \delta u^{(n)} - \frac{\dddot\phi}{\dot\phi} u^{(n)} = u^{(n-1)} \, .
\ee
The leading order solution $u^{(0)}$ is given by any linear combination of $u_{1,2}$. With boundary conditions $u^{(n)}(0) = \dot u^{(n)}(0)= 0$ when $n\geq1$, the higher order solutions are found iteratively from
\be
    u^{(n)}(t) = - u_1(t) \int^t_0 \td t'\, u_2(t') u^{(n-1)}(t') + u_2(t) \int^t_0 \td t'\, u_1(t') u^{(n-1)}(t').
\ee
Denoting by $u^{(n)}_i(t)$ the correction generated to $u^{(0)} = u_i$ and using the fact that the correction as well as its derivative vanishes at $t = 0$, we can expand $\Tr G = \sum_n (-k^2)\Tr G^{(n)}$.

The first correction is then
\bea\label{eq:k0growth_1st}
    \Tr G^{(1)}
&   = u_1(0) \dot{u}^{(1)}_2(T) - \dot{u}_1(0)u^{(1)}_2(T) + u^{(1)}_1(T) \dot{u}_2(0) - \dot{u}^{(1)}_1(T) u_2(0) \\
&   = (u_2(T)\dot u_2(0) - u_2(0)\dot u_2(T))\int^T_0 \td t'\, u_1(t')^2
\eea
We used the periodicity of $\phi$ and the conservation of the Wronskian at the $k^2$-order, $\det (w^{(0)} + w^{(1)}) = 1 + \mathcal{O}(k^4)$ which implies $u_1 \dot{u}^{(1)}_2 - \dot{u}_1 u^{(1)}_2 = u_2 \dot{u}^{(1)}_1 - \dot{u}_2 u^{(1)}_1$. Plugging in $u_1 = \dot \phi$, $u_2$ from Eq.~\eqref{eq:u2_piecewise} and using the definition \eqref{def:W}, we obtain
\be
    \Tr G^{(1)} = W \partial_\rho T \, .
\ee

\subsubsection*{Mode growth in the $k \to \infty$ limit}

For large $k$, the changes in the potential can be treated adiabatically. In particular, we require that $\dot\omega_k \ll \omega_k^2$, where $\omega_k = \sqrt{k^2 + U''}$. At leading order in $k$, this is equivalent to $|U''' \dot \phi| \ll k^3$. In this case, the two independent solutions of the mode equation \eqref{eq:deltaphi_eq_noH} are approximately 
\be
    u_{1,2} = \omega_k^{-1/2} \exp(\pm i \int \td t \, \omega_k) \, , 
\ee
so that
\be
    \frac{1}{2} \Tr G = \cos\left( \int^{T}_{0} \td t \, \omega_k \right) \, ,
\ee
since $\omega_k$ is periodic in $T$ when the potential is symmetric (or $T$ when it is asymmetric). As $|\Tr G| \leq 2$, we find that $\mu_k = 0$ confirming the well-known result that modes with an adiabatic evolution will not grow.

\section{Some exact results for the \texorpdfstring{$n=2$}{n=2} potential}
\label{app:n=2}

In $H \to 0$ limit, the field equation \eqref{FRW} admits an exact solution for the potential \eqref{eq:U} with $n=2$, \ie, when $U = U_0 \tanh^2(\phi/\phi_0)$. This solution reads
\be
    \phi(t) = \phi_0 \, {\rm asinh}\left[\frac{\cos(\pi t/T)}{\sqrt{U_0/\rho - 1}}\right] \, ,
\ee
and
\be
    W = \pi \sqrt{2U_0} \phi_0\left(1 - \sqrt{1 - \rho/U_0}\right), \qquad
    T = \frac{\pi \phi_0}{\sqrt{2(U_0 - \rho)}}
\ee
and thus satisfies \eqref{eq:W_expU} with $C=1$.

The two independent solutions to the $k=0$ mode equation read
\be
    u_1 = -\frac{\sin\left(\frac{\pi t}{T}\right)}{\sqrt{\frac{U_0}{\rho} - \sin^2\left(\frac{\pi t}{T}\right) }}, \qquad
    u_2 = -\frac{\frac{U_0}{\rho}\cos\left(\frac{\pi t}{T}\right) + \frac{\pi t}{T} \sin\left(\frac{\pi t}{T}\right)}{\sqrt{\frac{U_0}{\rho} - \sin^2\left(\frac{\pi t}{T}\right) }}\, .
\ee
In this exact solution, the linear growth of $u_2$ is explicit. From \eqref{eq:k0growth_1st} it directly follows, that
\be
    \frac{1}{2} \Tr G
    = - 1 - \frac{k^2 T^2}{2} \left( (1 - \rho/U_0)^{-1/2} - 1 \right)
\ee
consistent with \eqref{eq:mu_expU} when $C=1$.

\bibliography{main}

\end{document}